\documentclass[sigplan, screen, dvipsnames, pbalance]{acmart}
\usepackage{algorithmic}
\usepackage{graphicx}
\usepackage{textcomp}
\usepackage{tablefootnote}
\usepackage{fancyhdr}
\usepackage{siunitx}
\usepackage{colortbl} %
\usepackage{pifont} %
\usepackage{wasysym} %
\usepackage{marginnote} %
\usepackage[labelfont=bf, textfont=bf, aboveskip=5pt, belowskip=0pt, parskip=0pt]{caption,subfig}

\usepackage{setspace}

\usepackage[normalem]{ulem}

\usepackage{tcolorbox}
\usepackage{listings}
\usepackage{tabularx}
\usepackage{fancyvrb}
\usepackage[labelfont=bf, textfont=bf, aboveskip=5pt, belowskip=0pt, parskip=0pt]{caption,subfig}
\usepackage{enumitem}
\usepackage[italic]{mathastext}
\usepackage[subtle]{savetrees}

\usepackage{hyperref}
\usepackage{minted}

\definecolor{myOrange}{RGB}{255, 100, 120}

\newif\ifcameraready
\camerareadytrue
\ifcameraready
  \newcommand{\todo}[1][]{}
  \newcommand{\sg}[1]{{#1}}
  \newcommand{\rw}[1]{{#1}}
  \newcommand{\ad}[1]{{#1}}
  \newcommand{\ben}[1]{{#1}}
  \newcommand{\revsg}[1]{{#1}}
  \newcommand{\revRW}[1]{{#1}}
  \newcommand{\revad}[1]{{#1}}
  
  \newcommand{\crRW}[1]{{#1}}
  \newcommand{\crsg}[1]{{#1}}
\else
  \newcommand{\todo}[1][]{\textbf{\scriptsize \fcolorbox{black}{red}{\color{white}{TODO}}} \underline{$\overline{\hbox{\emph{#1}}}$}}
  \newcommand{\sg}[1]{\textcolor{black}{#1}}
  \newcommand{\rw}[1]{\textcolor{black}{#1}}
  \newcommand{\ad}[1]{\textcolor{orange}{#1}}
  \newcommand{\ben}[1]{\textcolor{black}{#1}}
  \newcommand{\revsg}[1]{\textcolor{blue}{#1}}
  \newcommand{\revRW}[1]{\textcolor{black}{#1}}
  \newcommand{\revad}[1]{\textcolor{orange}{#1}}
  
  \newcommand{\crRW}[1]{\textcolor{magenta}{#1}}
  \newcommand{\crsg}[1]{\textcolor{BrickRed}{#1}}
\fi

\newcommand{\uop}{\SI{}{\micro\nothing}op}
\newcommand{\uops}{\SI{}{\micro\nothing}ops}

\usepackage{tikz}
\newcommand{\ballnumber}[1]{%
  \tikz[baseline=(char.base)]{
  \node[shape=circle,draw,inner sep=0.5pt,fill=.,text=white,font=\small\bfseries] (char) {#1};}%
}

\renewcommand{\arraystretch}{0}
\DeclareSIUnit{\nothing}{\relax}

\AtBeginDocument{%
  }

\copyrightyear{2026}
\acmYear{2026}
\setcopyright{usgovmixed}
\acmConference[ASPLOS '26]{Proceedings of the 31st ACM International Conference on Architectural Support for Programming Languages and Operating Systems, Volume 2}{March 22--26, 2026}{Pittsburgh, PA, USA}
\acmBooktitle{Proceedings of the 31st ACM International Conference on Architectural Support for Programming Languages and Operating Systems, Volume 2 (ASPLOS '26), March 22--26, 2026, Pittsburgh, PA, USA}
\acmPrice{}
\acmDOI{10.1145/3779212.3790151}
\acmISBN{979-8-4007-2359-9/2026/03}

\begin{document}
\title{DARTH-PUM: A Hybrid Processing-Using-Memory Architecture}

\author{Ryan Wong}
\affiliation{
    \institution{Univ.\ of Illinois Urbana-Champaign}
    \country{Urbana, IL, USA}
}
\author{Ben Feinberg}
\affiliation{
    \institution{Sandia National Laboratories}
    \country{Albuquerque, NM, USA}
}
\author{Saugata Ghose}
\affiliation{
    \institution{Univ.\ of Illinois Urbana-Champaign}
    \country{Urbana, IL, USA}
}
\renewcommand{\shortauthors}{Ryan Wong, Ben Feinberg, \& Saugata Ghose}

\begin{abstract}
Analog \emph{processing-using-memory} (PUM; a.k.a. \emph{in-memory computing}) makes use of electrical interactions inside memory arrays to perform bulk \emph{matrix--vector multiplication} (MVM) operations.
However, many popular matrix-based kernels need to execute non-MVM operations, which analog PUM cannot directly perform.
To retain \crRW{its} energy efficiency, analog PUM architectures augment memory arrays with CMOS-based domain-specific fixed-function hardware to provide complete kernel functionality, but the difficulty of integrating such specialized CMOS logic with memory arrays has largely limited analog PUM to being an accelerator for machine learning inference, or for closely related kernels.
An opportunity exists to harness analog PUM for general-purpose computation: recent works have shown that memory arrays can also perform \crsg{Boolean} PUM operations, albeit with very different supporting hardware and electrical signals than analog PUM.

\revsg{We} propose \emph{DARTH-PUM}, a \revsg{general-purpose} hybrid PUM architecture that tackles key hardware and software challenges to integrating analog PUM and \crRW{digital} PUM.
We propose optimized peripheral circuitry, coordinating hardware to manage and interface between both types of PUM, an easy-to-use programming interface, and low-cost support for flexible data widths.
These design elements allow us to build a practical PUM architecture that can execute kernels fully in memory, and can scale easily to cater to domains ranging from embedded applications to large-scale data-driven computing.
\rw{We show how three popular applications \crsg{(AES encryption, convolutional neural networks, large language models)} can map to and benefit from DARTH-PUM, with speedups of \crRW{59.4$\times$}, \crRW{14.8$\times$}, and \crRW{40.8$\times$} over an analog+CPU baseline.}

\end{abstract}



\maketitle

\section{Introduction}
\label{sec:introduction}

\sg{Modern systems often suffer from the overheads of moving data between the CPU cores and the memory subsystem, with latencies and energy costs that are several orders of magnitude greater than those of computational instructions~\cite{Dally15, Boroumand2018.asplos, Oliveira2021}.}
\sg{These overheads are particularly harmful for} many modern and emerging workloads from machine learning~\cite{Gemini, Devlin2019.arxiv, Touvron2023.arxiv}, to databases~\cite{reinsel.whitepaper2018, Domo2023, Im2018}, to scientific computing~\cite{Kestor2013.iiswc, Pop2014.advhep, Pielke2013.ap, Scholkopf2004.mit}.
\sg{\emph{Processing-using-memory} (PUM; also known as in-memory computing or \emph{in situ} computing)} offers a potential solution to circumvent data movement overheads, by enabling intelligent memories 
\sg{that leverage} the electrical interactions of the memory devices to realize a meaningful computation.

PUM has attracted substantial interest in both academia~\cite{Li2016.dac, Truong2021.micro, Oliveira2024.hpca, Hajinazar2021.asplos, Zha2018.asplos} and industry~\cite{Khaddam2021.vlsis, Khaddam2022.jssc, mythic.m1076.website} due to its \sg{potential for large performance and energy improvements}.
\crsg{PUM can be classified as analog or \crRW{digital},} with both \crsg{types} having been explored in a variety of applications~\cite{Shafiee2016.isca, Sun2017.nvmsa, Zha2018.fpga, Zha2018.asplos, Feinberg2018.isca, Li2020.tcad, Li2024.hpca, Wong2025.isca} and memory device types~\cite{Feinberg2018.hpca, Aga2017.hpca, Seshadri2013.micro, Guo2011.micro, Kvatinsky2014.tcasii, Guo2013.isca, Gao2021.micro}.

Analog PUM is usually characterized by its ability to store multiple bits per cell and compute fast, though sometimes approximate, matrix\revad{--}vector multiplies by using a combination of Ohm's Law and Kirchhoff's Current Laws.
Although promising, analog PUM is susceptible to noise and other analog non-idealities; therefore, analog PUM has primarily been applied as a matrix-multiply accelerator for algorithms resilient to such errors (e.g., machine learning~\cite{Chi2016.isca, Song2017.hpca, Chou2019.micro, Bennettt2020.irps, Agrawal2022.iedm} and scientific computing~\cite{Feinberg2018.isca, Song2023.sc, Xiao2024.arxiv, Zuo2025.natelectron}).
\revsg{While analog PUM has spurred significant research and commercial prototypes~\cite{Khaddam2021.vlsis, Khaddam2022.jssc, mythic.m1076.website}, it requires expensive peripheral circuits (e.g., analog-to-digital and digital-to-analog converters), and either dedicated specialized function units or constant host--PUM communication to perform non-matrix operations required for machine learning and scientific computing.
\crsg{This highly constrains} analog PUM's applicability and benefits.}

In contrast, \crRW{digital PUM (i.e., Boolean PUM)} is often much more error-resilient and can perform \revsg{general-purpose} computations.
Notably, \crRW{digital} PUM uses single\revRW{-}bit cells and often realizes a Boolean operator between memory devices~\cite{Kvatinsky2014.tcasii, Seshadri2017.micro, Truong2022.jetcas}.
When executed sequentially, \revRW{these} sequences are able to \revRW{perform} \textit{any} computation, similar to a CPU.
Thus, \crRW{digital} PUM has been applied to a variety of data-intensive workloads~\cite{Oliveira2024.hpca, Hajinazar2021.asplos, Truong2021.micro}.
Despite having been \revRW{explored in} a variety of memory devices, including SRAM~\cite{Aga2017.hpca, Eckert2018.isca, Fujiki2019.isca, Wang2019.isscc, Wang2020.jssc, Khadem2025.hpca}, DRAM~\cite{Seshadri2013.micro, Seshadri2015.cal, Li2017.micro, Seshadri2017.micro, Gao2019.micro, Hajinazar2021.asplos, Gao2022.micro, Olgun2023.taco, Olgun2021.isca, Olgun2023.tcad, Yuksel2024.hpca, Oliveira2024.hpca, Yuksel2024.dsn, Jahshan2024.cal}, ReRAM~\cite{Kvatinsky2014.tcasii, Li2020.tcad, Truong2021.micro, Truong2022.jetcas}, MRAM~\cite{Guo2013.isca, Yan2016.nvmts, Angizi2018a.dac, Angizi2018b.dac, Angizi2019.dac}, PCM~\cite{Guo2011.micro, Cassinerio2013.advmaterials, Li2016.dac, Hoffer2022.jxcdc}, and NAND flash~\cite{Tseng2020.iedm, Tseng2021.vlsi, Shim2022.jetcas, Gao2021.micro, Park2022.micro, Chen2024a.micro, Kim2025.micro}, \crRW{digital} PUM often performs substantially worse compared to its analog counterparts when executing matrix-based applications.

\revsg{Prior works
have shown that analog PUM 
can be specialized for specific applications  (e.g., hyper-dimensional computing~\cite{Liu2025.tcad}, transformers~\cite{Li2024.hpca, Song2025.isca}) with the addition of \crRW{digital} PUM built to support application-specific functionality.}
Although these \revsg{works} leverage both domains of PUM computation, 
\revsg{they do not harness the generality of \crRW{digital} PUM in a way that can be extended to other applications}.
\revsg{Our goal in this work is to demonstrate that with careful design, a true hybrid PUM architecture can integrate both analog and \crRW{digital} PUM such that the same architecture can provide a best-of-both-worlds solution across many domains.}

\revsg{We} propose DARTH-PUM (\textit{\textbf{\underline{D}}igital--\textbf{\underline{A}}nalog \textbf{\underline{R}}esistive \textbf{\underline{T}}iles for \textbf{\underline{H}}ybrid \crRW{\textbf{\underline{P}}rocessing-\textbf{\underline{U}}sing-\textbf{\underline{M}}emory}}), which combines both analog and \crRW{digital} PUM onto the same chip.
\revsg{DARTH-PUM avoids the need for external (i.e., PUM--CPU) data movement or application-specific functional units, by taking advantage of the generalizable properties of \crRW{digital} PUM.}
We explore the design choices and functionality required to enable a hybrid PUM chip capable of executing a variety of applications without hardware redesign.
DARTH-PUM can execute a variety of applications (and datasets) by supporting auxiliary operations needed by analog \revRW{PUM} in nearby \crRW{digital} PUM units.
DARTH-PUM \revsg{provides flexible functionality}, with support for previously proposed techniques that optimize both analog (e.g., accuracy~\cite{Song2025.isca}, noise mitigation~\cite{Xiao2023.cas}, bit-slicing~\cite{Xiao2020.apr}) and \crRW{digital} (e.g., \crsg{vector computing~\cite{Aga2017.hpca},} bit-\crsg{pipelining}~\cite{Truong2021.micro}) PUM independently. 
Furthermore, DARTH-PUM supports a variety of bit width operands in both digital and analog PUM.

We make a series of key insights 
\crsg{about} flexible hybrid PUM.
(1)~\crsg{It} can substantially reduce  communication costs, as well as capacity overheads, of \crRW{digital} PUM. 
(2)~\crsg{It} can 
support a wide variety of bit-width operands
without the need for redesigning expensive post-processing circuitry;
furthermore, 
\crsg{digital PUM} can support all auxiliary operations needed by analog-PUM-accelerated kernels.
(3)~Bottlenecks with digital-to-analog conversions and vice-versa can originate from both analog and digital periphery.
\crsg{The} rate at \ad{which} analog and \crRW{digital} PUM arrays produce and consume data can directly affect the number of required peripheries (e.g., ADCs).
\crsg{We} analyze which functionalities are required for hybrid PUM (e.g., memory I/O), overheads (e.g., inter-tile communication) when executing these applications, and how to cleanly rate match the analog-to-digital conversions (and vice-versa). 

We demonstrate the performance benefits of DARTH-PUM and highlight its flexible nature by evaluating it on a collection of three different workloads: cryptography, \crsg{convolutional neural networks}, and large language models.
\revsg{Compared to an analog-only PUM accelerator coupled with a state-of-the-art CPU,} DARTH-PUM achieves a \crRW{59.4$\times$}, \revRW{14.8$\times$}, and \crRW{40.8$\times$} \revsg{performance} improvement while reducing energy by \crRW{39.6$\times$}, \crRW{51.2$\times$}, and \crRW{110.7$\times$} for the aforementioned workloads, respectively. 

\section{Background}
\label{sec:background}

\crsg{In this section, we provide a brief summary of background information related to analog PUM and digital PUM.}

\subsection{Processing-in-Memory}
Processing-in-memory (PIM) broadly encompasses the technique of bringing computation closer to the memory (or storage) substrate.
\revRW{\crsg{PIM} can be broadly divided into two \revsg{categories: \emph{proc\-es\-sing-near-memory} (PNM) and \emph{pro\-ces\-sing-using-memory}} (PUM).}
\crsg{PUM~\cite{Pawlowski2011.hcs, Nair2015.ibmjrd, Ke2020.isca,  Kim2021.hcs, Ke2022.ieeemicro, Kim2022.ieeemicro, Lee2024.cal, Rhyner2024.pact, Hyun2024.hpca} adds} compute capabilities to the memory by \sg{integrating conventional logic close to the memory arrays}.
In contrast, \crsg{PUM} enables computation in the memory by leveraging the electrical characteristics of the memory devices to realize meaningful computation.
This work focuses on 
the trade-offs associated with 
\crsg{an architecture capable of both analog and digital PUM.}

\subsection{Processing-Using-Memory}
Both analog and \crsg{digital} PUM have been explored \crsg{for a number of} application contexts~\cite{Sharad2013.dac, Li2015.dac, Hu2016.dac, Shafiee2016.isca, Xie2018.date, Feinberg2018.isca, Chen2024a.micro, Chen2024b.micro}, \sg{across a wide range of memory technologies~\cite{Aga2017.hpca, Seshadri2013.micro, Seshadri2017.micro, Kvatinsky2014.tvlsi, Guo2011.micro, Guo2013.isca, Xiao2022.tcas, Gao2021.micro}.
In this work, we use ReRAM because of its relatively mature support for both analog and digital operations.}

\subsubsection{Analog PUM}
\label{ssec:bkgd:analogpum}
Analog PUM is often characterized by its ability to store multiple bits per device \revRW{and has been explored as both an accelerator for \emph{matrix--vector multiplication} (MVM) and as an analog content-addressable memory (CAM; Section~\ref{para:related:analogcam}).
In this work, we focus on using analog PUM for efficient MVM, \revsg{as DARTH-PUM builds upon this functionality}.}

Computation \crsg{in analog PUM} \revsg{leverages} the shared bitline within the memory array \revRW{to execute add operations between all devices along the bitline in parallel.
Therefore,} the bitline can efficiently realize a multiply-and-accumulate (MAC) operation.\footnote{Analog PUM can also use single-level \crsg{devices (i.e., one bit per device), with the computation remaining a MAC.}}
Furthermore, by activating the entire analog array in parallel, full vector--matrix (or matrix--vector) multiplies can be realized.
Despite its promise as an efficient MVM accelerator, analog computation is susceptible to noise, limiting \sg{its} applicability.
Therefore, analog PUM has primarily been explored as part of application domains such as machine learning~\cite{Shafiee2016.isca, Chou2019.micro, Bojnordi2016.hpca, Hu2022.micro, Yuan2021.isca, Ji2019.asplos, Wang2021.iccad}, cloud analytics~\cite{Chen2024b.micro}, scientific computing~\cite{Feinberg2018.isca, Song2023.sc, Xiao2024.arxiv, Zuo2025.natelectron}, neuromorphic computing~\cite{Hu2016.dac}, and graph \revRW{applications}~\cite{Song2018.hpca, Liu2022.dac}.

\begin{figure}[h]
    \centering
    \includegraphics[width=\linewidth]{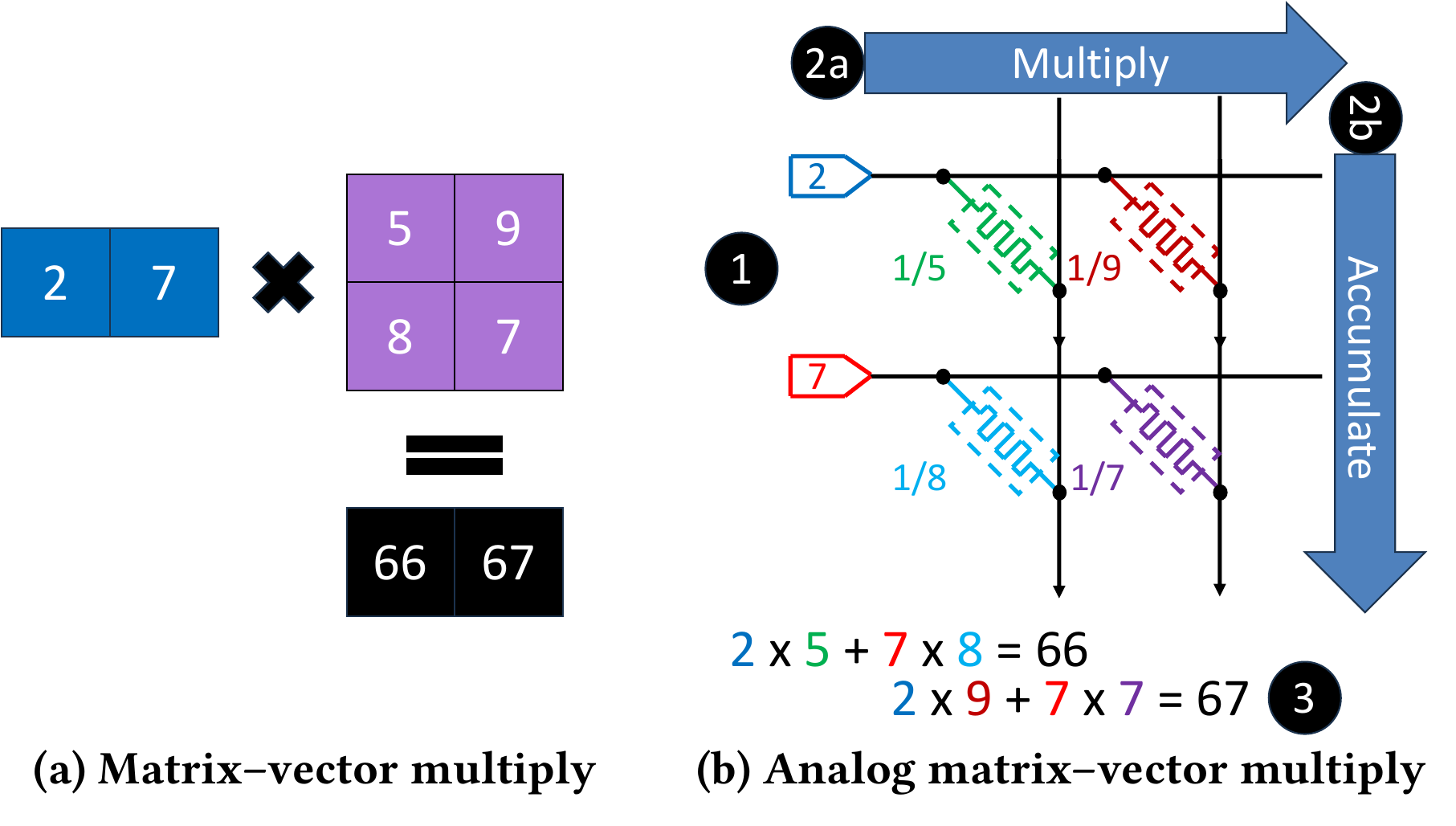}
    \caption{\crRW{Matrix--vector multiply with analog PUM.}}
    \label{fig:background:analog_PUM}
    \end{figure}

Figure~\ref{fig:background:analog_PUM} shows an example of a 2$\times$2 matrix multiplied by a 2$\times$1 input vector.
To execute the MVM, the matrix values are first programmed into the crossbar, usually in the form of resistances (or conductance, $G=\frac{1}{R}$).
\crRW{Each element of the input vector is then applied as a voltage to the wordlines (\ballnumber{1}).}
Using Ohm's Law ($I=\frac{V}{R}$), an element-wise \crsg{multiply} can be realized between the input and device \revRW{(\ballnumber{2a})}.
The current through each device is proportional to the input voltage times the device conductance \revRW{(\ballnumber{2b})}.
\ben{Similarly, by Kirchhoff's Current Law, the current through each device along the bitline is summed,}
\crsg{realizing the accumulation.}
\crsg{Finally, to} read out the value, analog PUM frequently employs the use of \crRW{analog-to-digital} converters (ADCs), which convert the analog current or voltages into \crsg{the discrete digital value of the MVM output} \revRW{(\ballnumber{3})}.

\begin{figure}
    \centering
    \includegraphics[width=\linewidth]{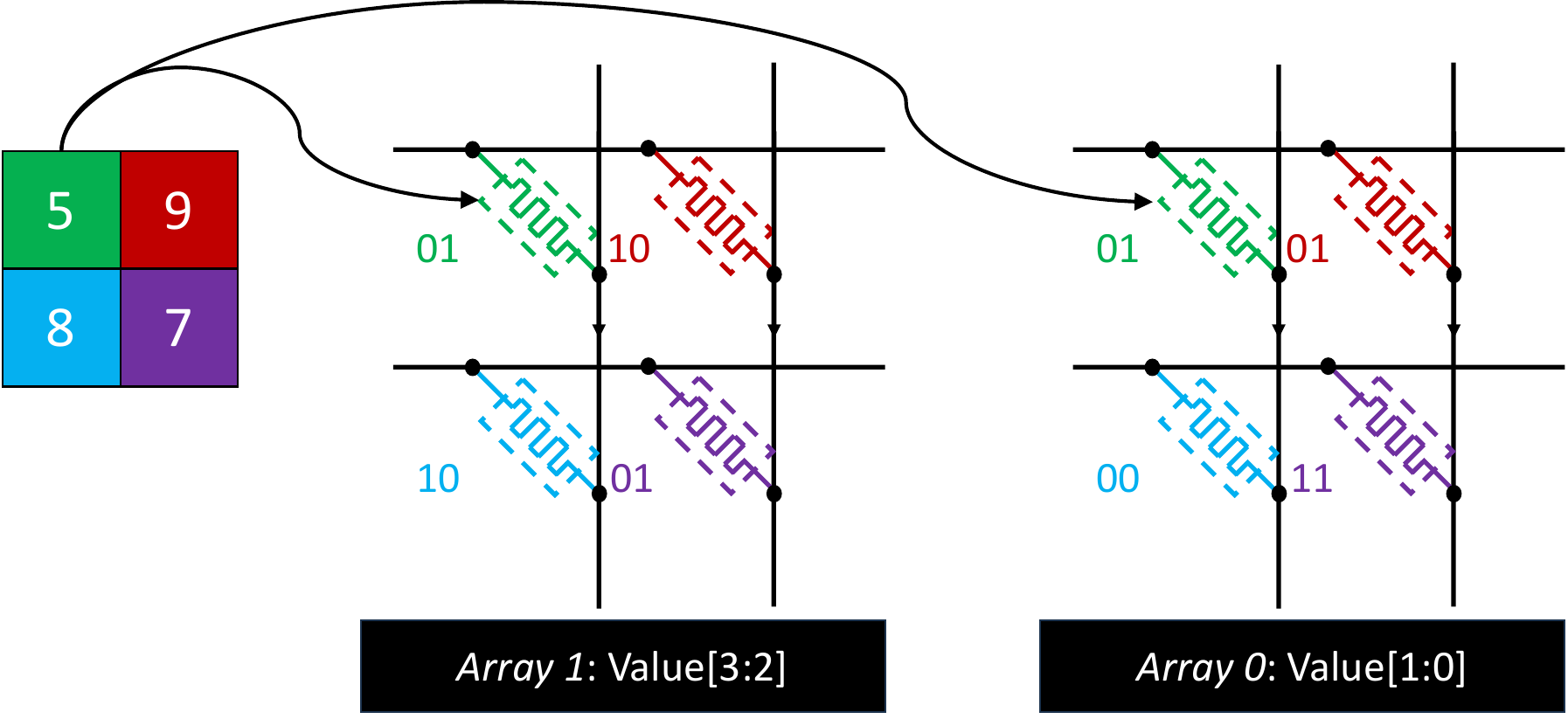}
    \caption{\crRW{Bit-slicing matrix values.}}
    \label{fig:background:bit_slice}
\end{figure}

\paragraph{Bit-Slicing.}
\ben{%
\crsg{For an analog PUM device, its effective maximum bit precision}
is approximately 6--12 bits, often constrained by the precision of the ADC used for the write--verify programming schemes.}
\crsg{To realize larger-width} operands, \sg{\emph{bit-slicing}} techniques~\cite{Shafiee2016.isca, Xiao2020.apr} 
split each stored \revRW{$N$-bit value} into smaller $\frac{N}{M}$ bit values, where $M$ is the number of bits that can \revRW{reliably} be stored in a single device.
Figure~\ref{fig:background:bit_slice} shows how a 4-bit value can be bit-sliced into two 2-bit values, which are then separated into different arrays (e.g., Array~1 stores \crRW{Value[3:2]}, while Array~0 stores \crRW{Value[1:0]}). %
\crsg{During MVM,} \revRW{each array executes the multiply} independently, resulting in multiple partial products.
To recombine the partial products into a single result, a post-processing step %
shifts the partial products based on the \revRW{relative} bit positions \revRW{of each bit slice}. 
In the example, the products from \revad{A}rray 1 must be shifted left by two bit-positions to account for bit-slicing.
\revRW{In contrast, the products from Array~0 are not shifted (i.e., shift by zero).}
The results are then added together forming the full output.

Bit-slicing can also be applied to input values, alleviating the need for high-range digital-to-analog converters (DACs).
An N-bit input can be sequentially applied, which results in $N$ partial products, which are handled during post-processing.
Notably, this recombination step is similar to the long-mul\-ti\-pli\-ca\-tion algorithm.
An in-depth discussion on bit-slicing for analog PUM can be found in prior work~\cite{Xiao2020.apr}.

\paragraph{Peripheral Circuitry}
\crsg{Peripheral} circuitry remains a substantial challenge for many analog PUM accelerators.
First, the array must be accompanied by costly \crsg{analog-to-digital} converters (ADCs), which convert the analog output current (or voltage) back into the discrete digital bit values.
Second, \crsg{digital-to-analog} converters (DACs) must be used to convert the digital inputs into analog input voltages.
Together, these periphery consume a non-trivial amount of energy and area, degrading the analog PUM efficiency. 
\crsg{Furthermore}, analog PUM requires additional logic for post-processing operations beyond the shift-and-add operation.
For example, additional logic, often in the form of \crsg{custom CMOS components}, must be added to the analog chip to execute auxiliary functions (e.g., activation functions in machine learning workloads), \crsg{reducing} the performance and flexibility of analog PUM.

\paragraph{Analog-to-Digital Converters}
For analog PUM, the type of ADC used can affect performance, power, and precision. 
\emph{Ramp} ADCs~\cite{Spear2023.jxcdc, Xiao2022.tcas} operate by comparing the input voltage with a predetermined reference voltage.
After each comparison, the reference voltage is increased \crsg{and compared again}.
By iterating linearly \crsg{over $N$~increments}, ramp ADCs take \crsg{$N$~cycles} to fully digitize the analog \crsg{input}.
\ben{%
\crsg{While ramp ADCs are often slower than other ADC types, they can} share the power-hungry reference generator\revad{,} increasing efficiency in applications where parallel readout is beneficial.}
\ben{\emph{Successive-approximation register} (SAR) ADCs~\cite{Kull2013.jssc} by contrast use binary search over the ADC range\revad{,} requiring $log(N)$ comparisons to find the correct output value.
\crsg{As an iterative binary search cannot be easily parallelized, SAR ADCs used for analog PUM} 
typically use a high-speed analog multiplexer to iterate over all the outputs.
In the context of bit-sliced systems, this approach has another advantage, since only a single value is output from the ADC at a time, thus only a single set of shift-and-add logic is required.}

\paragraph{Analog Non-Idealities.}
Analog devices \revad{are} susceptible to noise.
For example, many devices~\cite{Vontobel2009.nanotechnology, Xu2015.hpca} exhibit non-linear resistances (and therefore \crsg{non-linear} I--V curves), causing imprecision during programming.
Other non-idealities may affect device state over time (e.g., drift), and random perturbations on the bitline or device may affect the computation result.
Thus, analog PUM is often treated as an approximate solution and has primarily been applied to applications with inherent error tolerance (e.g., ML~\cite{Feinberg2018.hpca, Roy2019.nature}).
These non-idealities have been widely observed in a variety of prior works~\cite{Xiao2023.cas}. 

\begin{figure}[h]%
    \vspace{-15pt}%
    \hspace{3em}%
    \subfloat[Offset subtraction]{\includegraphics[width=0.32\linewidth]{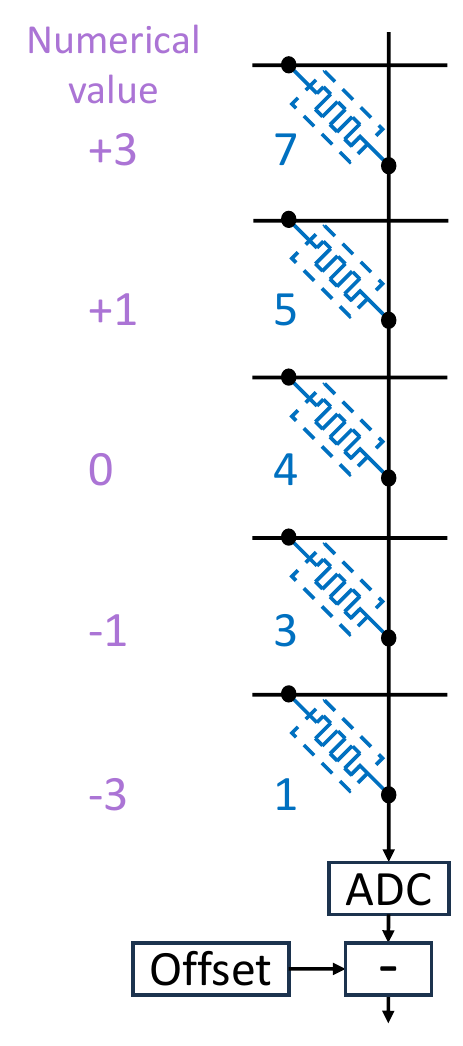}%
    \label{fig:background:negative_numbers-offsetsub}%
    }
    \subfloat[Differential cells]{\hspace{3em}%
    \includegraphics[width=0.25\linewidth]{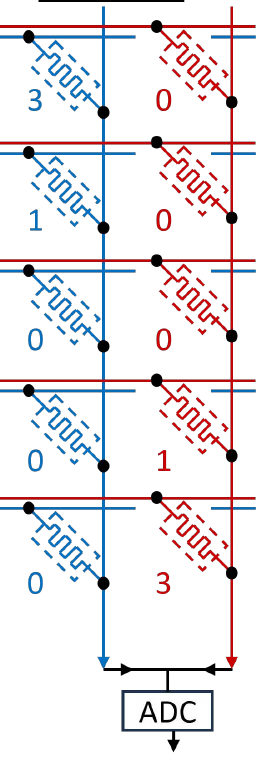}%
    \label{fig:background:negative_numbers-diffcells}%
    \hspace{3em}%
    }%
    \caption{\crRW{Number representations for the range [-3, +3]}.}
    \label{fig:background:negative_numbers}
    \vspace{-3pt}%
\end{figure}

\paragraph{\revRW{Handling Negative Numbers.}}
\label{para:bkgd:PUM:analog}
\revRW{
\revsg{Negative number representation} can be a challenge for analog arrays as resistance is a strictly positive quantity.
Two widely used \crsg{representations}
are offset subtraction~\cite{Shafiee2016.isca, Hu2016.dac, Hu2016.isqed, Nag2018.ieeemicro, Ankit2019.asplos, Yang2019.isca} and differential cell pairs~\cite{Chi2016.isca, Marinella2018.jetcas, Joshi2020.natcomm, Yao2020.nature, Guo2017.iedm}.
Offset subtraction (Figure~\ref{fig:background:negative_numbers-offsetsub}) \crsg{splits} the conductance range into two regions, wherein 
a conductance below the midpoint is a negative value, \revsg{while one} above the midpoint is positive.
To compute the final value, an offset is subtracted from the result of the ADC.
Differential cell pairs (Figure~\ref{fig:background:negative_numbers-diffcells}) utilize two discrete devices to represent negative and positive values while applying input voltages of opposite polarity.
Unlike offset subtraction, differential cell pairs do not require an offset scalar as the resulting current is directly proportional to the result.
This work utilizes differential pairs, which has been shown to have increased resilience against parasitic effects~\cite{Xiao2021.sst}. 
For a more detailed discussion on number representations, we defer to prior work~\cite{Xiao2023.cas}.}

\subsubsection{\crRW{Digital} PUM}
\label{ssec:bkgd:digitalpum}
\crsg{Digital (i.e., Boolean) PUM} has demonstrated potential to accelerate data-intensive applications~\cite{Seshadri2015.cal, Seshadri2017.micro, Truong2021.micro, Oliveira2024.hpca} due to its ability to implement \revRW{bulk bitwise operators} with minimal errors.
Two features that often separate \crRW{digital} and analog PUM are that \crRW{digital} PUM
(1)~often stores only a single bit per device, and 
(2)~operates by realizing a Boolean logic operator between two \revRW{input} cells, creating discrete outputs of `0' and `1.'
\crRW{Digital} PUM has been explored in a wide variety of contexts including 
SRAM~\cite{Aga2017.hpca, Eckert2018.isca, Fujiki2019.isca}, DRAM~\cite{Seshadri2017.micro, Hajinazar2021.asplos, Gao2019.micro, Li2017.micro, Oliveira2024.hpca}, ReRAM~\cite{Kvatinsky2014.tcasii, Truong2021.micro, Truong2022.jetcas}, MRAM~\cite{Guo2013.isca, Yan2016.nvmts}, PCM~\cite{Guo2011.micro, Cassinerio2013.advmaterials, Li2016.dac}, and NAND flash~\cite{Gao2021.micro, Park2022.micro}.
\crsg{Each digital PUM implementation chooses one of several}
\textit{logic families}~\cite{Kvatinsky2014.tcasii, Gupta2018.iccad, Truong2022.jetcas}, which correspond to the specific device, Boolean primitives, and corresponding voltages and operating parameters needed to realize the \crRW{digital} PUM \revRW{operations}.

\begin{figure}[h]
    \vspace{-8pt}%
    \centering
    \includegraphics[width=0.9\linewidth]{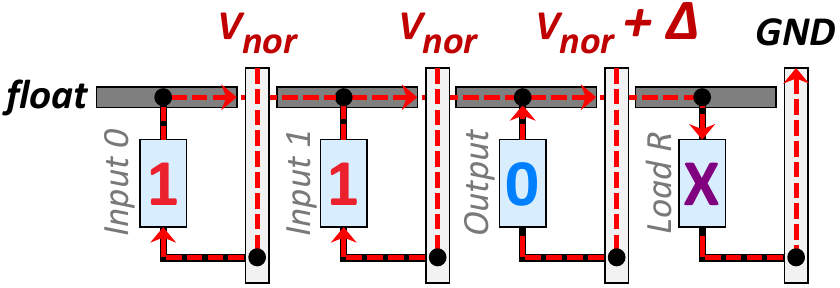}
    \caption{\crRW{Boolean} PUM operation with \crRW{OSCAR} \texttt{NOR}.}
    \label{fig:background:digital-pum}
\end{figure}

\crRW{One recently proposed logic family, OSCAR~\cite{Truong2022.jetcas}, implements the \crRW{\texttt{NOR}} and \crRW{\texttt{OR}} primitives using ReRAM devices.
Figure~\ref{fig:background:digital-pum} shows an example of the OSCAR \crRW{\texttt{NOR}} primitive, which uses four ReRAM devices as two inputs, an output, and a load resistor.
}
\crRW{To execute the \crRW{\texttt{NOR}} primitive, the two input bitlines are driven to $V_{NOR}$, the output bitline is driven to $V_{NOR} + \Delta$, and 
the load resistor bitline is set to GND \crsg{(with the resistor itself set between $R_{on}$ and $R_{off}$)}.}
The wordline is \revRW{then} electrically disconnected, leaving it in a floating state; thus, the wordline is perturbed by the interactions between the input cell states\crRW{, the output cell state, $V_{NOR}$, and $V_{NOR} + \Delta$}. 
Depending on the cell states, a current flows through the output device towards ground and \ad{changes} the state of the output \revRW{device}.\footnote{\crRW{By using a fourth resistor, OSCAR can properly balance the voltage division across the four devices, increasing fidelity of the operation.}}
To improve performance, entire columns of devices may be activated in parallel (with all wordlines in the \ad{array} floated), allowing for entire vectors of operands to simultaneously execute the \textit{same} primitive.

Beyond \crRW{OSCAR} \crsg{\texttt{NOR}}, prior works have realized several different Boolean operators (e.g., \texttt{AND}~\cite{Seshadri2015.cal, Park2022.micro}, \texttt{OR}~\cite{Seshadri2015.cal}, \texttt{NOR}~\cite{Kvatinsky2014.tcasii, Truong2022.jetcas}, \texttt{XOR}~\cite{Gupta2018.iccad}, \texttt{NOT}~\cite{Seshadri2017.micro, Aga2017.hpca}, \crRW{\texttt{IMPLY}~\cite{Kvatinsky2011.iccd, Kvatinsky2014.tvlsi})},
which provide different trade-offs and expressiveness.
Because \crsg{logic families are typically Boolean complete}, the %
PUM operations can be chained together to realize more complex operations (e.g., add, sub, cosine, square root).
\crRW{Digital} PUM often relies on bit-serial computation, wherein each bit of the output is computed sequentially, which \sg{significantly increases the latency of a single operation}~\cite{Truong2021.micro}.
Despite this \sg{overhead}, \crRW{digital} PUM can execute \textit{any} arbitrary function, enabling the same flexibility afforded to conventional CPUs.

\paragraph{Bit-Pipelining}

\begin{figure}
    \centering
    \includegraphics[width=\linewidth]{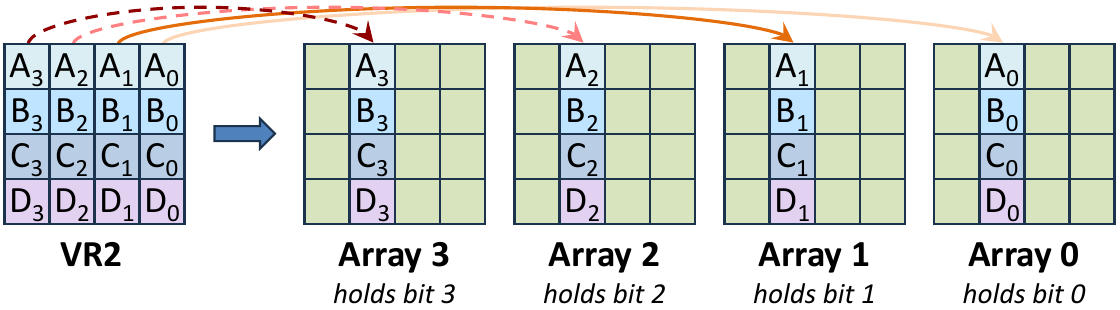}
    \caption{\crRW{Data layout for \crRW{digital} PUM pipeline.}}
    \label{fig:background:bit_pipeline}
\end{figure}

To improve throughput, RACER~\cite{Truong2021.micro} proposes \textit{bit-pipelining} for \crRW{\crsg{digital} PUM operations}.
In RACER, \revad{an} $N$-bit pipeline \revad{is} composed of $N$ discrete \crRW{digital} PUM arrays. 
Accordingly, an $N$-bit value is \textit{bit-striped} across all arrays (i.e., each array contains a single bit position of the value).
\revRW{Therefore, each \revad{array} in the pipeline can independently execute a different operation, enabling \revad{\textit{N}} times improvement in throughput.}
Figure~\ref{fig:background:bit_pipeline} shows how data is \revRW{striped} in a 4-bit pipeline.
Each \textit{vector register} (VR) contains four \crRW{\textit{elements} (\crsg{A--D in our example})}, which are distributed to different \textit{rows} of the pipeline.
Each element is bit-striped across the arrays (e.g., \crsg{for all elements in VR2, bit~0 is placed in column~2 of array~0}).
\revRW{Because operands are striped across different \revad{arrays} in the same row, different rows may contain different values; therefore, a} pipeline with $M$~rows can execute an \revad{$M$}-element vector in parallel.
Due to design constraints (e.g., limited peripheral circuitry), vector register elements can only interact with other bits along the same row (e.g., VR$_{0}$ element\crsg{~B can only execute a Boolean primitive with VR$_{N}$ element~B}).
Although bit-pipelining can substantially improve the performance for many data-parallel applications, this technique alone leaves substantial performance opportunities compared to analog PUM for matrix-based algorithms.

\section{Motivation: \revsg{The Potential of} \revRW{Hybrid} PUM}
\label{sec:motivation}

The limited flexibility of analog PUM has led researchers to primarily focus on application- \revsg{or domain-}specific accelerators, designed and tuned with specific peripherals \revsg{and specialized function units (SFUs).}
In contrast, \crRW{digital} PUM has 
\revsg{enabled a wide range of general-purpose accelerators}, but leaves a large gap in performance for MVM-style kernels. 

\begin{figure}[h]
    \vspace{-10pt}
    \subfloat[\revsg{\crRW{Digital} PUM}]{\includegraphics[width=0.32\linewidth]{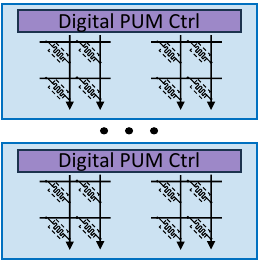}%
    \label{fig:motivation:digital-pum}%
    }
    \hfill%
    \subfloat[\revsg{Analog PUM}]{\includegraphics[width=0.32\linewidth]{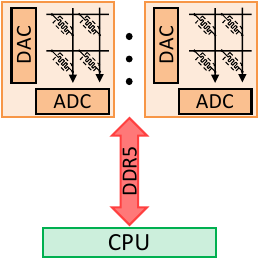}%
    \label{fig:motivation:analog-pum}%
    }%
    \hfill%
    \subfloat[\revsg{\crRW{Naive} Hybrid PUM}]{\includegraphics[width=0.32\linewidth]{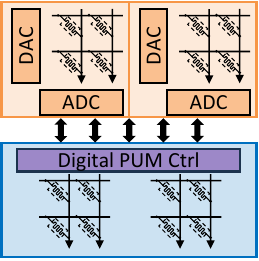}%
    \label{fig:motivation:hybrid-pum}%
    }%
    \caption{\revsg{Motivating architectures.}}
    \label{fig:motivation:arch}
\end{figure}

\revsg{Hybrid} PUM can address these issues by \revsg{combining} the complementary behavior of analog and \crRW{digital} PUM, \revsg{without the need for SFUs}.
\revsg{However, as} we \revsg{show in this section}, simply combining two high-performance schemes is insufficient to realize the full potential of \revRW{hybrid} PUM\ad{,} \revRW{as combining the two domains requires careful balancing and coordination between them}.
\revRW{To \revsg{illustrate why}, we \revsg{compare} three different architectures, \revsg{as} shown in Figure~\ref{fig:motivation:arch}:
\revsg{(1)~\crRW{digital} PUM (\crRW{D}), a RACER pipelined microarchitecture~\cite{Truong2021.micro} that performs \crRW{\texttt{NOR}} operations using the OSCAR logic family~\cite{Truong2022.jetcas} in ReRAM arrays;
(2)~analog PUM (A), \revRW{\revsg{a ReRAM-based version of} Xiao et al.~\cite{Xiao2022.tcas},} where non-MVM operations are performed by a \SI{4}{\giga\hertz} 8-core Arm CPU with 256-bit vector extensions; and 
(3)~nine configurations of a \crRW{naively combined} hybrid PUM (H), each with different \crRW{digital} PUM and analog PUM array counts, where \crRW{digital} PUM performs non-MVM operations.
These comparisons are iso-area to the Arm CPU (except for A, whose analog PUM area is treated as free).}}

\begin{figure}[h]
    \centering
    \vspace{-5pt}
    \includegraphics[width=\columnwidth]{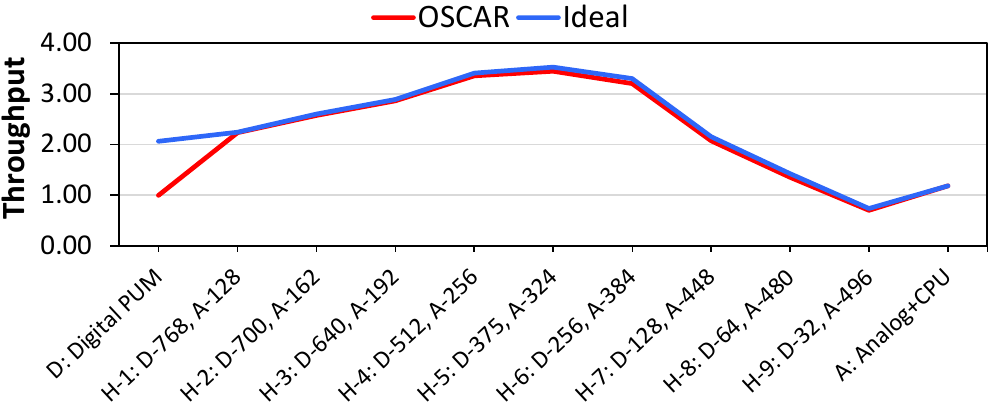}
    \caption{\revsg{Throughput of AES-128 encryption with \crRW{digital (D)}, analog (A), and \crRW{naive} hybrid (H) PUM with the OSCAR and ideal logic families, \revsg{normalized to D with OSCAR.}}}
    \label{fig:motivation:aes}
\end{figure}

\revsg{We explore the shortcomings of these architectures using AES (the Advanced Encryption Standard), a widely used block cipher~\cite{Nechvatal2001.nist, FIPS197}.\footnote{For a detailed explanation of AES, see Section~\ref{ssec:applications:aes}.}}
To encrypt data, a \SI{16}{\byte} input (plaintext) is organized as a 4$\times$4 matrix. %
The \revRW{AES} algorithm consists of four main steps:
(1)~\crRW{\texttt{SubBytes}}, which replaces each byte of the plaintext block with a byte from the substitution matrix (\crRW{S-box});
(2)~\crRW{\texttt{ShiftRows}}, which cyclically \revsg{left-}shifts each row based on a predefined value;
(3)~\crRW{\texttt{MixColumns}}, which is a matrix multiply between the plaintext and a predefined matrix; and 
(4)~\crRW{\texttt{AddRoundKey}}, which executes \revsg{an} \crRW{\texttt{XOR}} operation between the generated key and the plaintext.
\revsg{Figure~\ref{fig:motivation:aes} shows the iso-area throughput of AES-128 encryption for our three architectures.
We make three observations from the figure.}

\revsg{First, analog PUM performs only 18\% better than \crRW{digital} PUM, even though AES involves many MVMs.
This is because three of the four AES steps cannot be performed using analog PUM, and must be offloaded to the CPU.
To analyze this further, we study the CPU performance of AES using the gem5 simulator~\cite{gem5}, and find that non-MVM operations consume a majority of the execution time (even when including the data-movement overheads), being bottlenecked by the limited parallelism in the CPU compared to the analog PUM chip.}

\revsg{Second, most hybrid PUM configurations outperform both our pure \crRW{digital} PUM and pure analog PUM configurations.
We note that as we start from a fully \crRW{digital} PUM chip and convert some of the area into analog PUM arrays, the performance increases up to a peak point (configuration H-5), as the inclusion of analog PUM significantly accelerates the bottleneck step (\crRW{\texttt{MixColumns}}).
Beyond the peak, as more \crRW{digital} PUM area is converted to analog PUM, fewer \crRW{digital} PUM arrays are available, shifting the bottleneck to the inability to perform enough plaintext encryptions in parallel.
At its peak, hybrid PUM performs \revRW{3.54$\times$} better than \crRW{digital} PUM.}

\revsg{Third, the choice of logic family has minimal impact on most hybrid PUM configurations.
We introduce an \emph{ideal} logic family, capable of performing any two-input Boolean operation within a single cycle, to observe if a \crRW{\texttt{NOR}}-only architecture is a limiting factor.
This builds upon architectures such as FELIX~\cite{Gupta2018.iccad}, which show benefits for \crRW{digital} PUM by diversifying the number of available Boolean operators.
While the pure \crRW{digital} PUM configuration sees a $2.1\times$ improvement in throughput, due to significant reductions in executing the \crRW{\texttt{MixColumns}} MVM operations, this still falls short of a hybrid PUM configuration with just a small number of analog arrays and only OSCAR support.
In the best case for hybrid PUM, an ideal logic family increases throughput over OSCAR by only 3.2\%.
This means that we can build a hybrid PUM architecture without needing to support multiple Boolean operators, for which each additional operator increases the peripheral circuit area and complexity~\cite{Gupta2018.iccad, Truong2022.jetcas}.}

\revsg{From our analysis, we make two conclusions:
(1)~hybrid PUM can offer significant performance benefits by offering the best of both \crRW{digital} PUM and analog PUM in an iso-area footprint; and
(2)~hybrid PUM does not need complex logic families or SFUs to achieve its performance, improving its memory density and significantly simplifying the chip design.
However, we note that for our \crRW{naive} hybrid architecture to be successful, we would have to do a similar sweep of configurations for each separate architecture.
In Section~\ref{sec:system}, we show how we can avoid such customization by instead introducing co-designed hardware--software support to design and optimize a practical multi-domain hybrid PUM architecture.}

\section{\revsg{DARTH-PUM} System Design}
\label{sec:system}

\label{ssec:system:DARTH-PUM}
\begin{figure}[t]
    \centering
    \includegraphics[width=\linewidth]{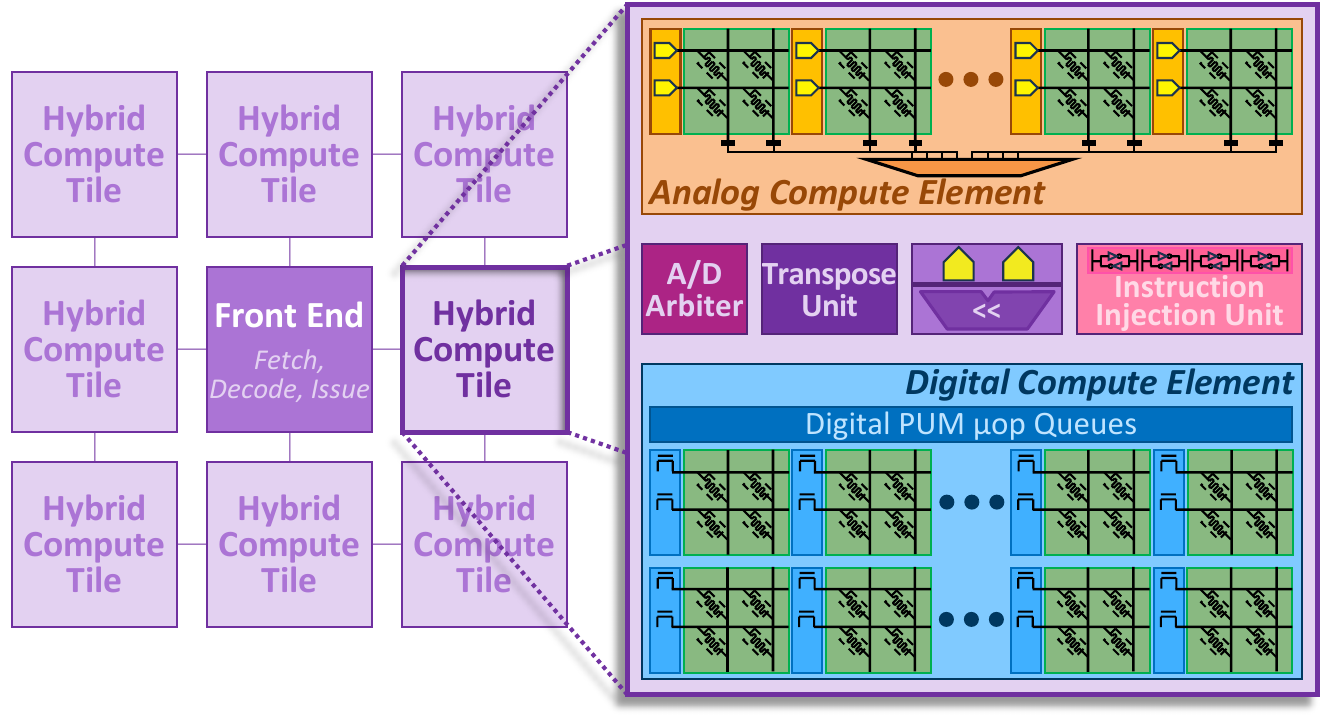}
    \caption{DARTH-PUM architecture.}
    \label{fig:system:darth-pum}
\end{figure}

\revRW{\revsg{Based on our conclusions from Section~\ref{sec:motivation}, we propose DARTH-PUM, a fixed architecture capable of providing best-of-both-worlds (i.e., \crRW{digital} PUM and analog PUM) benefits across many application domains.
DARTH-PUM consists of two parts, as shown in Figure~\ref{fig:system:darth-pum}:
(1)~a front-end controller, which can fetch and decode hybrid PUM instructions (Section~\ref{ssec:system:software}) and issue PUM-specific \uops{}; and
(2)~a back end composed of multiple \emph{hybrid compute tiles} (HCTs).}}

\paragraph{\revsg{Hybrid Compute Tile.}}
\revRW{Each HCT \revsg{contains}
(1)~an \textit{analog compute element} (ACE), 
(2)~a \textit{digital compute element (DCE)}, and 
\revsg{(3)~auxiliary components 
that manage the flow of \uops{} and optimize interactions between the ACE and DCE.}
\revsg{An ACE contains} 64 analog arrays and peripheral circuitry required for computation.
\revsg{A DCE contains} 64 \revsg{RACER-based} \crRW{digital} PUM pipelines, \revsg{provisioned} to provide storage and sufficient output pipelines for analog MVMs, and \revsg{RACER control circuitry that includes a digital issue queue to dispatch \uops{} to digital arrays within each pipeline}.
We connect the ACE and DCE using a\revad{n} I/O data transfer network capable of transferring \SI{8}{\byte} per cycle, which is \revsg{chosen} to \revsg{rate-match analog-to-digital converter (ADC) throughput} with \revsg{DCE write bandwidth}.}

\revsg{We introduce four types of auxiliary components in an HCT, as shown in Figure~\ref{fig:system:darth-pum}, to coordinate kernel execution and address issues with the \crRW{naive} hybrid PUM architecture from Section~\ref{sec:motivation}.
First, we introduce an analog--digital interface that consists of shift units and ADCs that carefully balance ACE and DCE throughput (Section~\ref{sec:system:interface}).
Second, 
we introduce an analog--digital arbiter and an instruction injection unit to sequence operations and data movement between the ACE and DCE (Section~\ref{sec:system:control}).
Third, we introduce an error compensation scheme specialized for the HCT (Section~\ref{ssec:system:parasitics}).
Fourth, we expose a novel software interface 
for
DARTH-PUM (Section~\ref{ssec:system:software}), with per-HCT instruction injection units to reduce the overhead of \uop{} expansion and mitigate front-end stalls.}

\begin{figure*}
    \centering
    \includegraphics[width=\linewidth]{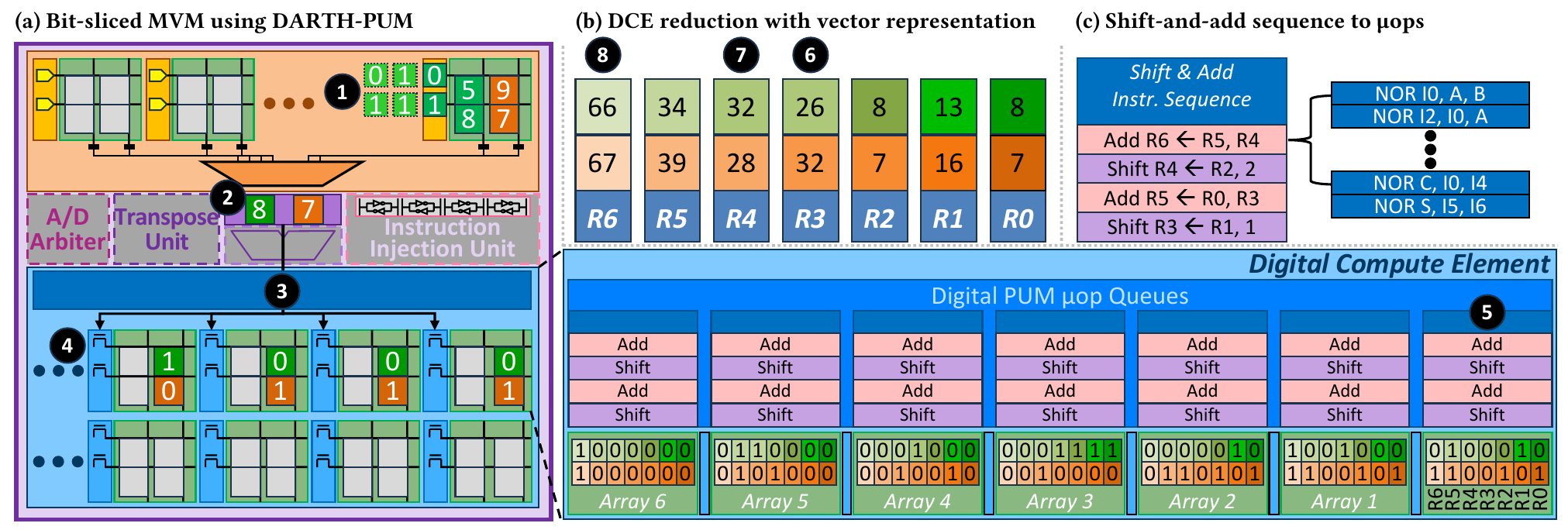}\vspace{4pt}
    \caption{\crRW{Unoptimized matrix--vector multiply with DARTH-PUM using ACE and DCE (\crsg{gray} components are unused).}}
    \label{fig:system:mvm-demo}
\end{figure*}

\paragraph{\revsg{Matrix--Vector Multiplication Walkthrough.}}
\revRW{%
Figure~\ref{fig:system:mvm-demo} shows how a 2$\times$1 vector is multiplied with a 2$\times$2 matrix using \revsg{the analog side of DARTH-PUM,} with \revsg{three-bit} inputs and \revsg{four-bit} matrix elements.
\revsg{The} matrix \revsg{resides} inside the \revsg{ACE},
\revsg{and the input vector is bit-sliced and placed \revsg{at} the ACE's input buffers (\ballnumber{1} in Figure~\ref{fig:system:mvm-demo})}.
\revsg{At} each cycle, a new bit of the input is applied to each wordline (horizontal), \revsg{performing a one-bit} MVM, and generating a vector of partial products \revsg{along each bitline (\ballnumber{2})}. %
The resulting partial products are then transferred from the ACE to the DCE via the on-chip network and bit-striped across different arrays within a DCE pipeline (\ballnumber{3}).
Each element of the partial product vector corresponds to a different element in the \revad{final }output vector; therefore, each element is stored on a separate row within the same DCE vector register (\ballnumber{4}).
As the ACE computes the remainder of the input bit positions, shift and add instructions are interleaved and added to the Digital Issue Queues (\ballnumber{5}).}

\crRW{As the result of each bit-sliced input is generated by the ACE and placed in the DCE, shift and add operations are executed to reduce the partial products.
For clarity, we show the registers without the bit-striped values (Figure~\ref{fig:system:mvm-demo}b), though the data remains in bit-striped format.
After each bit slice is processed by the ACE, the partial product is shifted by its corresponding bit position using digital \uops{}, \revsg{e.g.,} the partial products of bit position 1 (stored in register~R1) must be shifted once \revsg{into R3} (\ballnumber{6}).
Similarly, the result of bit position 2 (R2) must be shifted twice into R4 (\ballnumber{7}).
Between each shift, addition instructions accumulate a partial sum.
Once all partial sums have been accumulated, the final output vector is stored in R6 (\ballnumber{8}).
}

\subsection{\revsg{Interfacing the ACE With the DCE}}
\label{sec:system:interface}

To support input bit-slicing, \revsg{analog PUM architectures typically introduce dedicated shifter and adder SFUs that} can quickly recombine the partial products \revsg{at the cost of significant chip area}.
\revsg{Since our \crRW{unoptimized} hybrid PUM lacks explicit shifters and adders, the data must be written, shifted, and added using \crRW{Boolean} PUM operations.}
This creates a new performance bottleneck, as all three operations \revsg{cannot interfere} with one another.
Specifically, partial products cannot be added until they have been shifted into the proper bit position.
Simultaneously, the analog arrays are producing new partial products that must be written into the \crRW{digital} PUM \ad{arrays}.

\begin{figure}[h!]
    \centering
    \includegraphics[width=\linewidth]{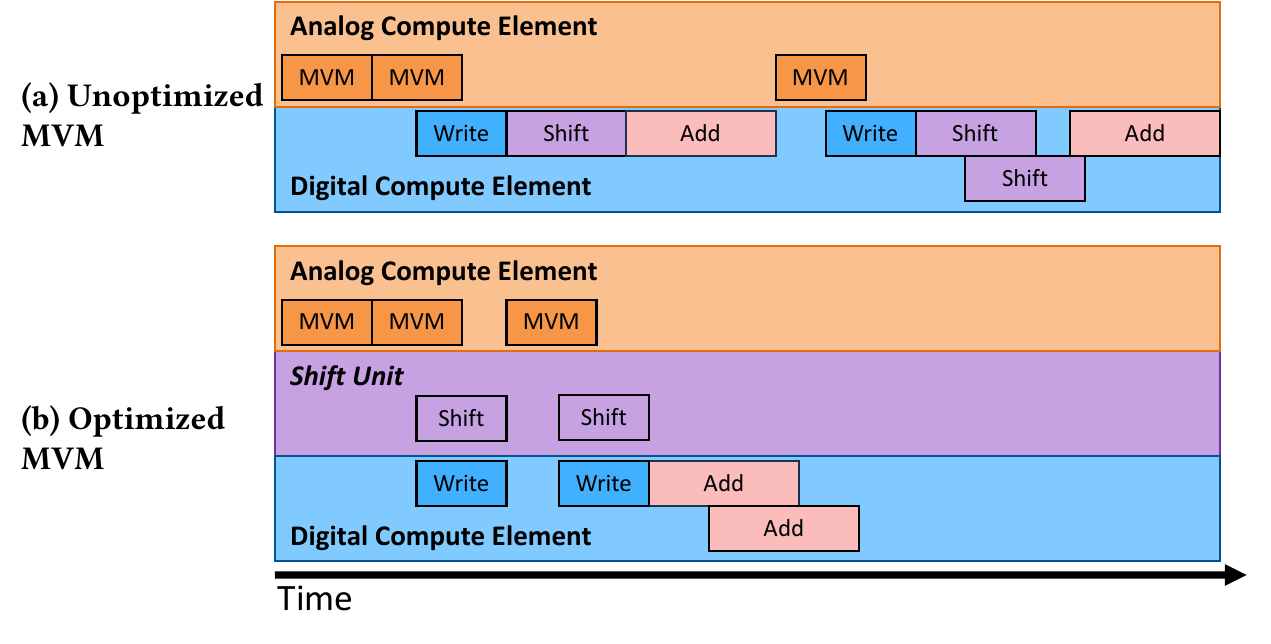}
    \vspace{-10pt}%
    \caption{\crRW{Sequence of operations for MVM.}}
    \label{fig:system:mvm-timeline}
\end{figure}

\revRW{Figure~\ref{fig:system:mvm-timeline}a shows the timeline of the unoptimized MVM shown in \crRW{Figure~\ref{fig:system:mvm-demo}}.
\crRW{The inputs are bit-sliced.
Thus, the ACE executes three discrete MVM operations} (orange \revsg{operations in \crRW{Figures~\ref{fig:system:mvm-demo}a} and \ref{fig:system:mvm-timeline}}), \revsg{producing three separate outputs that must be written into the DCE}.
However, \revsg{the DCE's} pipelines can only write one row of data per cycle. Thus, for a pipeline with $N$~elements per vector register, it would take $N$~cycles to write the partial products to the digital pipeline.
Furthermore, it would take another $M$~cycles to shift the data into the proper position (blue \revsg{operations}).
\revsg{This is due to each input bit~$i$ needing to be shifted $i$~bit positions to the left}. %
Therefore, for a pipeline with a width of $N$ and a depth of $M$, it would require $N$~cycles to write followed by $M$~cycles to shift the data before the next set of partial products can be written to the digital arrays.
Furthermore, the \crRW{digital} PUM would then need to execute bit-pipelined \crRW{\texttt{ADD}} operations prior to reading the next set of data (red \revsg{operations} in Figure~\ref{fig:system:mvm-timeline}).
Importantly, these operations (write, shift, add) cannot be interleaved or pipelined without disruptions to \revsg{the DCE's} bit-pipelining \revsg{execution model}, enforcing serialization.}

To alleviate this bottleneck \revsg{in DARTH-PUM}, we implement a series of shift units that can place data into the correct bit positions during each transfer from \revsg{the ACE to the DCE}, and wait until the \revsg{transfer completes before executing} the add operations \revsg{in the DCE}.
Because the shift amount can be predetermined based on the number of bits applied (e.g., 1-bit inputs require \revsg{a single} shift),
the shift units can write the data into the \revsg{DCE in a fixed manner without the need for software intervention or reconfigurable interconnects}.
By shifting during the data transfer, we gain three major benefits.
First, waiting \revsg{for} all partial products can avoid paying the warmup and cooldown costs of a shift operation per partial product. 
Second, the rate at which ADCs produce data can be matched with the rate at which \revsg{the DCE} can write data to \revsg{its} arrays.
Third, all \revRW{outstanding} \crRW{\texttt{ADD}} operations \textit{can} be pipelined, leveraging the benefits of digital pipelined PUM.
\revsg{Figure~\ref{fig:system:mvm-timeline}b shows the resulting timeline after optimization.}

\paragraph{Balancing Analog and Digital Array Counts}
In addition to balancing the rate at which partial products are produced and processed, 
\revsg{we need to carefully decide how to allocate the number of analog arrays in the ACE, and the number of digital arrays in the DCE.}
To optimize the balance between analog and digital arrays, sufficient arrays should first be allocated to hold the matrices (or sub-matrices) needed for the computation.
Notably, this can mitigate the need for writing \revsg{to} the multi-bit devices during computation, as storing data in analog devices can incur substantial latency overheads~\cite{Hu2016.dac, Ankit2019.asplos}.
Therefore, underallocating analog arrays may cause disruptions during the computation.

As a second consideration, \revRW{the number of analog and digital arrays} also influences the amount of peripheral circuitry required for analog computation.
Specifically, when only few arrays contain data required at any point in the computation (e.g., a single layer of a neural network), fewer \revsg{peripheral circuits} are required as fewer arrays are active at any one point in time (i.e., arrays containing later layers may be inactive while processing the earlier layers of the network).

The amount of parallel computation executed during the application may also influence the type of \revsg{peripheral circuits required} (e.g., SAR ADC vs.\ ramp ADC).
In the case of the AES algorithm, the matrix contains many rows, but few columns; 
therefore, an ideal system would be able to convert many bitline outputs in parallel. 
\revsg{While a ramp ADC consumes greater area than an SAR ADC, it requires less power; we evaluate the impact of ADC choice in Section~\ref{ssec:eval:adcs}.}

The outputs produced by the analog arrays are effectively wide-width row vectors (where each element is generated as a bitline output).
To recombine \revsg{these into a single value}, the vectors must be added using \revsg{the DCE}.
Therefore, the \revsg{DCE} can be treated as an SIMD vector unit, where the number of elements is equal to the number of rows.
To account for this, sufficient bandwidth as well as capacity must be allocated in \revsg{the DCE}. 
For example, increasing the number of active arrays will increase the number of digital pipelines required to store and process the output data.
Similarly, fewer active \crRW{digital} PUM arrays may enable \revad{the architecture to tolerate} \crRW{longer} ADC latencies or fewer ADC units, as fewer outputs can be processed per cycle.

\subsection{\revsg{Coordinating ACE and DCE Operations}}
\label{sec:system:control}

Whereas analog-only PUM can be easily enabled via a series of library calls (e.g., fixed-function fixed-width MVM), \revRW{hybrid} PUM requires a more sophisticated interface. 
Our goal with \revRW{hybrid} PUM is to deploy entire applications to the device.
Therefore, we propose a full ISA to execute analog and digital operations \revRW{and \revad{an} accompanying library (Section~\ref{ssec:system:software})}.
We observe that digital ISA instructions can be made to only interact with digital arrays; however, analog-type instructions require coordination between \textit{both} analog and digital arrays.
Therefore, we develop instructions to coordinate which can be used to prevent analog and digital operations from interfering \revsg{with one another}.

With our proposed ISA, instructions can be dispatched to analog and digital arrays from a single instruction stream.
However, this creates two problems, which we describe below,
(1)~analog and digital instruction interference,
and
(2)~\crRW{contention for the digital PUM pipelines due} to shift-and-add.

\paragraph{Mitigating Instruction Interference}
First, analog and digital instructions have different latencies and dependencies.
Notably, analog instructions may take hundreds of cycles to execute (due to ADC overheads and I/O from the array); in contrast, digital operations \crRW{may} consume only \crRW{\crsg{up to} tens of} cycles (in the case of a Boolean primitive). 
\crRW{Thus,} even though the analog instruction may consume hundreds of cycles (e.g., MVM), the entire MVM should appear atomic \crRW{(i.e., new digital instructions should not interleave with the reduction sequence in Figure~\ref{fig:system:mvm-demo}c)}, without disruption from digital instructions.

Therefore, to manage the instruction sequence and prevent instructions from colliding, we provide an arbiter in hardware, which limits an array to either analog or digital operations.
Importantly, this also provides serialization, ensuring that a younger instruction cannot execute until an older instruction has completed.
Therefore, a digital instruction that has dependence on a prior analog instruction (e.g., ReLU after an MVM) will be stalled until after the analog MVM has completed.

\rw{Over the course of the analog matrix multiply, numerous partial products are generated and placed in the digital vector-registers.
This creates a potential for \crRW{existing data to be overwritten} in \revRW{hybrid} PUM.
Recall that partial products are shifted and placed into the digital arrays before any products are summated.
Thus, these products may require up to N temporary vector registers for an $N$-bit input.
Without a mechanism to determine which vector registers are available or unavailable, the analog arrays may overwrite data inside ``live'' vector registers.
Unfortunately, dynamically determining which vector registers may and may not be used to to store intermediate values could add substantial cost.}
To prevent data corruption, we introduce a \textit{pipeline reserve} instruction \revsg{for DARTH-PUM}, which allocates a pipeline, effectively marking all data stored in the pipeline as ``dead.''
Notably, this data can be efficiently moved to another pipeline prior to executing the MVM instruction, and this constraint can easily be enforced by either a compiler or runtime system.

\paragraph{Reducing Issue Logic Stalls}
In addition to the shift-and-add, the add sequence may consume hundreds of additional micro-ops to combine the partial results.
Recall that \crRW{digital} PUM implements Boolean operations; therefore, a single \crRW{\texttt{ADD}} instruction may be decomposed into tens of Boolean operators \crRW{(Figure~\ref{fig:system:mvm-demo}c)}. 
Therefore, relying on the \crRW{issue} logic to recombine the partial products would consume substantial bandwidth and cause the issue and dispatch logic to stall on every MVM until all partial products have been generated and combined.
To mitigate \ad{these} overheads, we introduce \revsg{a hardware} \rw{\emph{instruction injection unit} (IIU)} in \revsg{DARTH-PUM}.
We leverage the observation that the same instruction, \crRW{\texttt{ADD}}, is executed sequentially and repeatedly but with the only change being the vector register arguments.
Thus, the IIU can be implemented as a small table and a counter, which is able to issue the Boolean operation sequence directly to the digital pipelines.
By adding the IIU, \revsg{DARTH-PUM} can free up \revsg{its front-end issue logic to issue instructions to other HCTs}.

\paragraph{Moving Data Between \revsg{the ACE and the DCE}}
Even with optimizations of allocations, applications require shuffling of data between analog and digital arrays; 
however, the operands in the two discrete systems operate on different \revRW{ax\revad{e}s}.
Specifically, analog PUM applies a row-wise (wordline) input, but combines in a columnar direction (bitline). 
In contrast, \crRW{digital} PUM stripes data column-wise, where each piece of data is a different bitline, but computes row-wise (e.g., activating two different bitlines along the same row produces a new bit in a new bitline). 
Therefore, data that crosses between analog and digital requires transpositions.

To accelerate this, we introduce a \revsg{\emph{transposition unit}}, which assists during both MVM computation and data storage.
First, when input bit-slicing is enabled, the output of the analog array contains a vector of partial products.
To combine the subsequent partial products (e.g., prod$_i$ + prod$_{i+1}$), the vector must be shifted and subsequently added to the next output vector.
Importantly, each \textit{element} of the output vector~$i$ must interact with the \ad{corresponding} element of vector~$i+1$.
\revsg{To} place the outputs correctly in the digital arrays, the analog \textit{row vector} must be transposed into a \textit{column vector} (i.e., placed into a vector register).
Notably, this transposition must occur for each partial product. 
Second, when data \revsg{(e.g., a copy of a matrix)} \ad{is} 
\revsg{moved from digital to analog or vice-versa, that} data must be transposed. 
When a matrix is stored in the digital arrays, each vector register (column) is allocated to a column of the matrix. \crsg{When} the matrix is programmed into an analog array, each row contains the data stored in the digital vector register.

\paragraph{Supporting Non-Vector Operations}
\revRW{RACER only provides support for vector copy operations, thereby incurring high overheads when only specific elements of a vector register are required.}
To retrieve a singular element, the \crRW{digital} PUM must first copy a vector, apply a\ad{n} element-wise mask, and execute an \crRW{\texttt{AND}} operation. 
Furthermore, because the \crRW{digital} PUM is pipelined, the \crRW{\texttt{AND}} must be applied sequentially to every bit before the next operation can proceed.
This overhead can be prohibitively expensive in applications that require loading only a single element based on an address specified in the vector register (e.g., AES \crRW{\texttt{SubBytes}}). 
Notably, this would require the entire vector copy and \crRW{\texttt{AND}} sequence for every element in the vector register (e.g., 64\crRW{$\times$} vector \crRW{copy} + 64\crRW{$\times$} \crRW{\texttt{AND} operations} for a 64-element vector register).

To alleviate this, we implement element-wise loads and stores, which sequentially fetches the address specified by each element in the register. 
On each cycle, a row of \crRW{digital} PUM is read out, which is used as an address.
\rw{The \crRW{digital} PUM can then read the data from an adjacent pipeline (again in one cycle) and transfer it back to the original pipeline.
In the case of AES, other pipelines may be pre-loaded with the \crRW{S-box} (e.g., for \crRW{\texttt{SubBytes}}) or other plaintexts to encrypt. For other applications (e.g., ML), these pipelines may store weight matrices for layers not actively in use.
To reduce complexity, we limit the address range of the load and store to only digital pipelines within the same Hybrid Compute \revRW{Tile} (see Section~\ref{ssec:system:DARTH-PUM}).}

\paragraph{Expanding to Large-Width Operands}
Analog accelerators usually target specific bit-width operands (e.g., 8b), and design their post-processing digital logic with these outputs in mind;
however, given the flexible nature of \crRW{digital} PUM, a \revRW{hybrid} system can be dynamically configured to handle both different bit width operands and \crRW{support analog devices with a different number of bits per cell.}
For example, an application may elect to use 8b operands, which are distributed across 2$\times$4b arrays. 
Alternatively, the same application may require higher-precision, thereby limiting the devices to 2 bits per cell, requiring 4$\times$2b arrays.
Notably, the only change in post-processing is the shift lengths and arguments needed for the \crRW{\texttt{ADD}} operation.

To support variable bit-widths, we introduce the idea of \revsg{a \textit{virtual analog core} (vACore)}.
\revRW{A \revsg{vACore} logically combines multiple analog arrays within a single analog compute element together to support \revad{wide-width operands}.
}
Additionally, allocating a \revsg{vACore} configures the shift units as well as the instruction injection unit to transparently implement the shift-and-add sequence.
The \revsg{vACores} can be tracked through firmware and can be changed when the application requests either a different bit width, or the number of bits per cell stored in the arrays. 
Using the \revsg{vACore} abstraction, applications can dynamically adjust precision (based on the number of bits per cell) and support larger matrix values.
\revRW{Note that to reduce complexity, the HCT can only have \revsg{vACore}\revad{s of the same bit width} at a time.}

\subsection{Compensating for Errors}
\label{ssec:system:parasitics}
\revRW{
As noted in Section~\ref{ssec:bkgd:analogpum}, analog MVMs are susceptible to noise.
Therefore, for accuracy-critical applications, errors during the computation in the analog domain can lead to incorrect results.
For example, during AES encryption, an error during the \crRW{\texttt{MixColumns}} operation can change the ciphertext in a way that prevents decryption.
In our analysis, we observe that using the analog PUM arrays in SLC mode is insufficient to ensure an error-free approach.
To mitigate these concerns, we develop a parasitic compensation scheme to mitigate against analog non-idealities.
}

\begin{figure}[h]
    \centering
    \includegraphics[width=\linewidth]{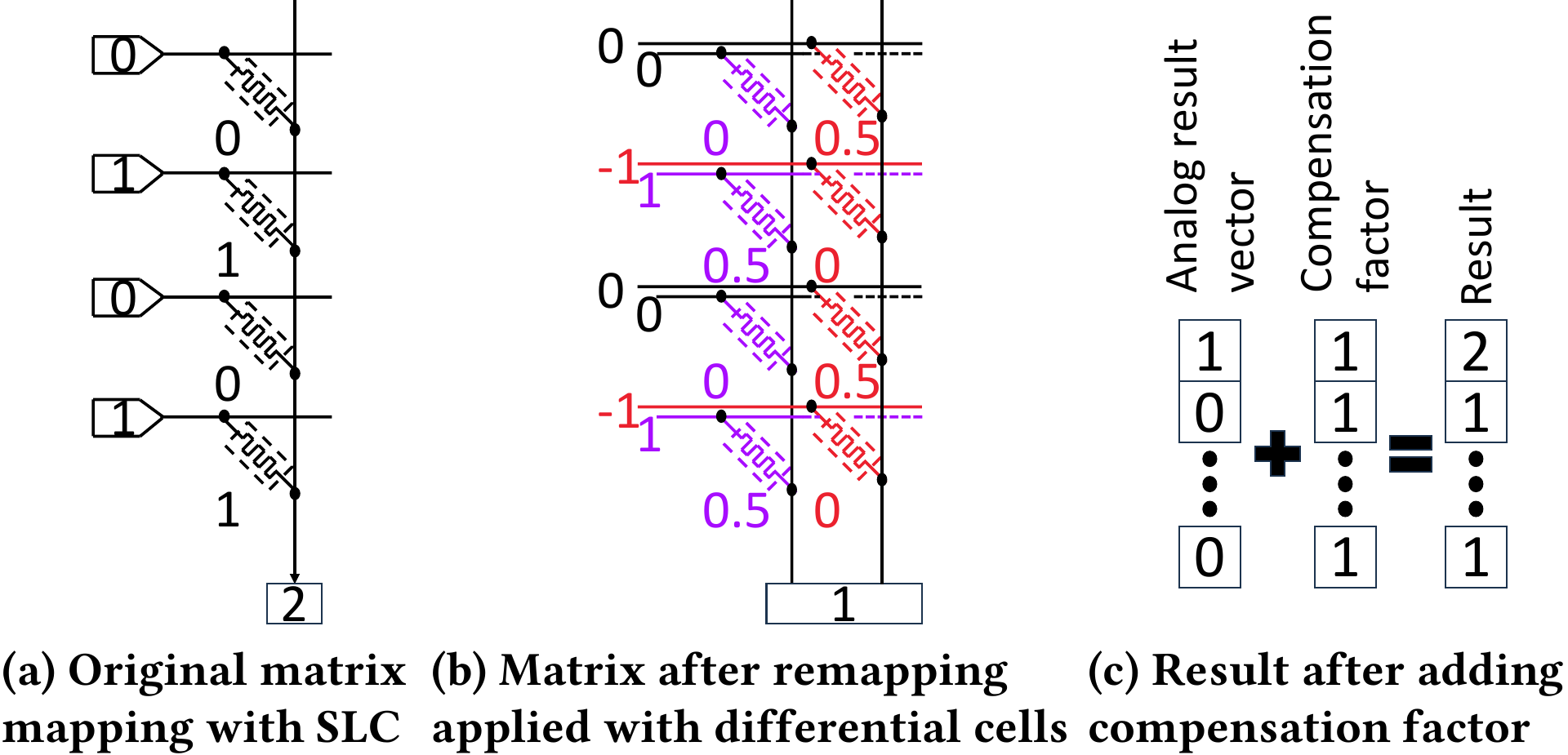}
    \caption{\revRW{Steps for parasitic compensation scheme.}}
    \label{fig:parasitics}
\end{figure}

\revRW{
Our parasitic compensation scheme consists of two components:
(1)~a remapping technique, in which the matrix values are stored \revsg{using alternative representations;} and
(2)~a post-processing error-correcting code (ECC) compensation factor, which is applied using \revsg{the DCE}.
Figure~\ref{fig:parasitics} shows \revsg{the original values, data remappings}, and 
compensation factor.
}

\revRW{
Our remapping technique leverages the observation that the AES matrix is strictly positive.
This means that each device along a bitline contains \revsg{a bit value of} 0 or 1.
Recall that we use differential cell pairs (Section~\ref{para:bkgd:PUM:analog}); therefore, because the matrix is entirely positive, the device storing the negative value is always storing a 0.
Thus, the current down each column becomes relatively large  causing large IR \revsg{(current/resistance; i.e., Ohmic)} drops along the positive bitline~\cite{Xiao2021.sst}.
To mitigate these behaviors, we remap the \revsg{bit values 0/1 to -1/+1}, respectively, using differential cell pairs~\cite{Xiao2023.cas}.
By remapping values from strictly positive to both positive and negative, the currents down each bitline are reduced, which brings the effect of IR drop below one ADC \revsg{least significant bit} level \revsg{(i.e., it will no longer cause the ADC to read out an incorrect value)}.}

\revRW{
To further reduce the effects on non-idealities, we leverage the sparsity of the matrix and input vector, as well as the fixed number of inputs, to reduce the magnitude of the error.
Specifically, we combine the observation that the input vectors are sparse and contain a fixed number of 1s.
For example, AES has a total of four 1s in the input vector.
Additionally, we can scale the range of the matrix values from [-1, 1] to [-0.5, +0.5], effectively reducing the computed value by half.
By reducing the range of the values, we can simultaneously reduce the magnitude of the noise.
To re-scale the output to the correct value, a post-MVM factor can be applied to the result of the MVM.
In the case of AES, a scale factor of 2 is applied as an addition during post-post processing (4$\times$0.5).\footnote{\crRW{Figure~\ref{fig:parasitics} uses a factor of 1 (2$\times$0.5).}}
Importantly, this post-processing is applied to the entire output vector (i.e., all output values of the \revsg{ACE}).
Note that this operation can be efficiently realized using a vector operation in the nearby \revsg{DCE}, which can then be used to execute subsequent PUM operations.
Our parasitic compensation scheme leverages algorithm insights to effectively mitigate noise by leveraging both analog and \crRW{digital} PUM, further highlighting the benefits of hybrid PUM.
}

\revRW{
As noted in Section~\ref{para:bkgd:PUM:analog}, differential cell pairs are not exclusive to AES and can be broadly applied to all applications \revsg{of} DARTH-PUM.
To generalize, the remapping technique leverages prior work in parasitic compensation~\cite{Xiao2021.sst, Xiao2022.tcas}, which highlights the effects of current cancellation.
Similarly, the parasitic compensation factor can be extrapolated by knowing the relative sparsity of the input vector and the relation the stored matrix.
}

\subsection{Software Support}
\label{ssec:system:software}
\revRW{
Tracking the internal state of HCTs, including parameters, across the entire chip can provide a burdensome task for programmers.
Therefore, to ease programmer complexity when using DARTH-PUM, we provide both a library and a full hybrid ISA.
The library is comprised of two types of commands: application-agnostic and application-specific.
Using these library API calls, the explicit handling of \ad{\revsg{vACores}} can be made transparent to the programmer.
}

\begin{table}[t!]
\centering
\caption{List of \revsg{DARTH-PUM library calls.}}
\label{tab:system:software:commands}
{\footnotesize
\setstretch{0.7}
\renewcommand{\arraystretch}{2}
\begin{tabularx}{\columnwidth}{l X}
\toprule
\textbf{API Call} & \textbf{Description} \\

\arrayrulecolor{black}\midrule
\multicolumn{2}{c}{\textit{Application-Agnostic Calls}} \\
\arrayrulecolor{black}\midrule
\texttt{allocVACore()} & Allocates a vACore on an HCT, based on  \textit{elementSize} and bits per cell. Initializes the HCTs with the shift-and-add \uop{} sequence. \\
\texttt{setMatrix()} & Allocate and store a matrix with the required number of HCTs. \\
\texttt{execMVM()} & Execute MVM between matrix and vector. \\
\texttt{updateCol()} & Updates a column of a matrix on HCTs. \\
\texttt{updateRow()} & Updates a row of a matrix on HCTs. \\
\texttt{disableAnalogMode()} & Disables ACE and copies matrix from analog to digital arrays. \\
\texttt{disableDigitalMode()} & Disables DCE post-processing. \\

\arrayrulecolor{black}\midrule
\multicolumn{2}{c}{\textit{Application-Specific Calls}} \\
\arrayrulecolor{black}\midrule
\texttt{AES\_initArrays()} & Reserve HCTs. Copies S-box to \crRW{DCE}. Sets the matrix needed for \crRW{\texttt{MixColumns}} in the analog arrays.\\
\texttt{AES\_<en/de>crypt()} & Encrypt/decrypt. Optional argument for input key.\\
\texttt{CNN\_setModel()} & Allocates and stores the model layers to HCTs. Mapping is done as a per-layer distribution.\\
\texttt{CNN\_runInference()} & Executes inference operation.\\
\texttt{CNN\_changeActivation()} & Changes the activation function used between layers during inference.\\
\texttt{LLM\_build<En/De>coder()} & Allocates and stores encoder/decoder on HCTs.\\
\texttt{LLM\_runInference()} & Execute inference operation.\\
\texttt{LLM\_changeActivation()} & Changes the activation function used between layers during inference.\\

\arrayrulecolor{black}\bottomrule
\end{tabularx}
}%
\end{table}

\paragraph{\revRW{Application-Agnostic Commands}}
\revRW{
Application-agnostic commands provide general-purpose MVM support and transparently manage the \crsg{ACE--DCE} dataflow 
without \crsg{making any} assumptions on the intended workloads.
These commands can be used by programmers with minimal a priori knowledge of the underlying hardware (e.g., that it has analog and \crRW{digital} PUM).
As seen in Table~\ref{tab:system:software:commands}, these commands enable programmers to \crRW{allocate and} interface with the stored matrices \crRW{in} DARTH-PUM.
For example, to store a matrix in the HCT, a user would use the \revsg{\texttt{setMatrix()} operation, which takes three arguments:} the matrix, \crRW{\textit{elementSize} (which refers to the bit width of each matrix element)}, and the level of bit precision required by the programmer.
To reduce complexity, we elect to have bit precision be a scale (e.g., 0--2), which we map to 1 bit per device, \revsg{half of all possible bits per device, and the maximum possible number of} bits per device.
For an 8b device, this would correspond to 1, 4, and 8 bits per cell, respectively.
Using this information, the runtime system can allocate the number of HCTs required to store the matrix \emph{without any more programmer input}.}

\paragraph{\revRW{Application-Specific Commands}}
\revRW{
For our evaluated workloads, we develop application-specific commands better tailored to high-level APIs.
We develop these commands for users with minimal to no knowledge of the underlying hardware.
Therefore, these API calls are designed to closely model high-level functions, while including sufficient meta-information to enable the system to map the computation to DARTH-PUM.
For example, the \texttt{setModel()} command used for CNNs can build the DARTH-PUM mapping by allocating HCTs on a per-layer basis.
\texttt{setModel()}-specific layer types (e.g., fully connected, convolution) are mapped to analog arrays within the HCT, while leaving other computations (e.g., bias, activations) in the \crRW{digital} PUM.
Additionally, to elide the complexities of analog PUM and precision, the bit-precision parameter \revad{is abstracted} as an ``accuracy'' target, which is exposed as a scale to the programmer.
We design the accuracy parameter to mirror the performance--accuracy trade-off between \crRW{storing} more bits per cell and fewer HCTs, versus fewer bits per cell and more HCTs, which affects total throughput.}

\paragraph{\revRW{Assisting Expert Programmers}}
\revRW{
To support new applications and enable greater flexibility for expert programmers, we include a full hybrid ISA. %
Additionally, we provide an additional set of commands to directly manage the \revsg{vACores} on the HCT, enabling (1)~the ability to allocate the \revsg{vACores} on an HCT, 
(2)~the precision at which they are operating, and 
\revsg{(3)~updates to the data stored} inside the HCT.
Note that these commands are hardware-centric, enabling the programmer to directly change the configuration of the HCT, whereas the application-agnostic commands are matrix-centric, enabling programmers to work at both the individual \revsg{array or software levels}.
The combination of the ISA and application-agnostic commands provides developers with the flexibility to create their own custom mappings from applications to DARTH-PUM.
Similarly, these \revsg{functions} enable expert programmers to leverage the wealth of prior work on analog and digital mapping techniques~\cite{Shafiee2016.isca, Hu2016.dac, Feinberg2018.isca, Ankit2019.asplos, Xiao2023.cas, Song2025.isca} to improve different metrics (e.g., throughput, precision, latency, peripheries).
For even greater flexibility, we provide options to enable or disable the \revsg{ACE or DCE within each HCT}, allowing developers to \revsg{more precisely customize} where computation is performed.}

\section{Applications}
\label{sec:applications}
We describe how DARTH-PUM can be used to improve the performance of three different applications, \crsg{convolutional} neural networks, LLM encoders, and AES encryption.

\subsection{\crsg{Convolutional} Neural Networks}
\crsg{Convolutional} neural networks (CNN) have been widely explored as part of the analog PUM literature.
Although various analog accelerators exist~\cite{Shafiee2016.isca, Xiao2022.tcas, Chi2016.isca}, they are often engineered for a specific set of functions (e.g., activation functions) or neural network topology.
\revRW{Of the main operations needed for CNN execution (convolution, fully connected \revsg{or FC} layers, downsampling, pooling, activation functions, and batch normalizations), only \revsg{convolution and FC layers can make use of analog MVM}.}
Therefore, \revsg{each layer requires some degree of} extra digital logic \revsg{(i.e., specialized function units)}, in some cases~\cite{Shafiee2016.isca, Nag2018.ieeemicro, He2020.micro} requiring periphery for dedicated shift-and-add network, in addition to the expensive analog periphery.

\revsg{In contrast, with DARTH-PUM, the entire CNN pipeline can be executed on a single chip without the need for SFUs.}
We place the layers of the neural network (matrices) into the analog arrays, while reserving digital pipelines for input vectors.
Convolution layers leverage a Toeplitz expansion~\cite{Strang1986.sam} that maximizes the number of rows.
To handle layers which exceed the size of the arrays, we decompose the MVM into a series of MVMs, which are issued sequentially. 
Once the operations have completed, including the shift-and-add, the results are combined in the digital pipelines by using vector COPY instructions from different pipelines followed by ADD operations.
We leverage digital PUM to implement the remaining auxiliary functions that analog PUM would have needed external support for (e.g., pooling, activation functions).
For increased efficiency, inputs can be batched by storing multiple sets of inputs in the inactive digital pipelines.
This presents a cost trade-off, as indefinitely increasing the batch size may affect performance due to the numerous in-flight operations.

\subsection{LLM Encoders (Transformers)}
LLM encoders are widely used in modern ML applications~\cite{Devlin2019.arxiv, Topal2021.arxiv} (e.g., LLMs).
Specifically, these encoders rely on attention mechanisms~\cite{Vaswani2017.nips}, which require softmax, square root, and layer normalization. 
These operations cannot be accelerated using only analog PUM.
Therefore, in \revsg{place of} dedicated digital logic, DARTH-PUM relies on 
\revsg{its DCE} to realize the non-MVM operations using IBERT~\cite{Kim2021.pmlr} algorithms.

\revsg{Unfortunately, not all of the encoder's} MVM operations can be easily mapped to \revsg{DARTH-PUM's ACE}.
Specifically, the matrices used in the attention mechanism rely on dynamic updates, which can create disruptions, as analog PUM requires more time and energy to program compared to digital \crRW{PUM}\crRW{,} potentially leading to long overheads.
Therefore, we execute the computations needed by the attention mechanism in \revsg{the DCE}, while executing the feed-forward \crRW{network} \crRW(FFN) using \revsg{the ACE}\crRW{,} without the need for additional \revsg{SFUs}.

\subsection{Advanced Encryption Standard}
\label{ssec:applications:aes}
{As noted in Section~\ref{sec:motivation}, the AES algorithm encrypts a \SI{16}{\byte} plaintext using \revad{successive rounds of} four main functions, \crRW{\texttt{SubBytes}, \texttt{ShiftRows}, \texttt{MixColumns}, and \texttt{AddRoundKey}, which begin after an \texttt{AddRoundKey} intialization step.
In the last (10th) round, only \crRW{\texttt{SubBytes}, \texttt{ShiftRows}, and \texttt{AddRoundKey}} are executed.
AES-128 executes 10 rounds (\crsg{for a} 128-bit key), while AES-192 and AES-256 execute 12 and 14 rounds, respectively.
Figure~\ref{fig:applications:aes_layout} shows how each step of AES is mapped to DARTH-PUM. 
\crRW{\texttt{SubBytes}}} (\ballnumber{1}),
\crRW{\texttt{ShiftRows}} (\ballnumber{2}), and 
\crRW{\texttt{AddRoundKey} (\ballnumber{4}) are all executed using the \revsg{DCE. The}
\crRW{\texttt{MixColumns}}} step (\ballnumber{3}) leverages the \ad{ACE} for efficient execution.

During \crRW{\texttt{SubBytes}}, \crRW{each byte of the plaintext is replaced with a byte from the S-box}. 
Therefore, we store the \crRW{S-box} in an unusued pipeline \revRW{within each \revsg{HCT}}.
\crRW{Because the requested \revad{bytes} may not be contained in a single vector register, the DCE uses the new element-wise load instruction \revad{(Section \ref{sec:system:control})} to access the S-box pipeline data. }

\crRW{\texttt{ShiftRows}} requires a cyclic left shift depending on \revad{the} row of the block plaintext (e.g., row 1 is left shifted 1 full byte).
We implement this using pipelined left shifts; however, because there is no left terminal buffer, which could be used to shift the most-significant bit to the least-significant bit, we implement pipeline reversal and right shift operation. 
As part of this pipeline reversal macro, the entire digital pipeline is \crRW{drained}, followed by shift operations propagating in reverse mode. 
Note that we must fully drain the pipeline to preserve data integrity (e.g., reversing the pipeline while instructions are still present in the issue logic may cause only a subset of the bits to have the instruction applied).

\begin{figure}
    \centering
    \includegraphics[width=0.68\linewidth]{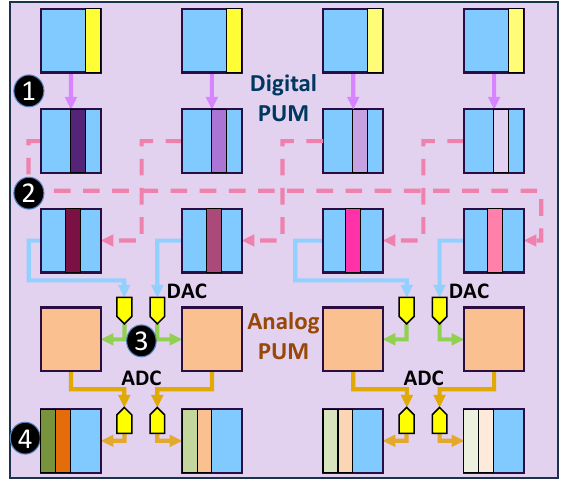}
    \caption{\crRW{Dataflow for AES encryption.}}
    \label{fig:applications:aes_layout}
\end{figure}

\crRW{\texttt{MixColumns}} can be implemented as an MVM between the input and a predetermined matrix, followed by a modulus operator over a Galois field~\cite{Carlitz1932.ajm}.
\crRW{Alternatively, lookup tables may be used, where the indexes are denoted by a vector.}
\crRW{Post-lookup (or MVM), an \crRW{\texttt{XOR}} is applied to the resulting bits to compute the final result.}
We use this insight to \crRW{perform} the \crRW{\texttt{MixColumns}} step in the analog domain by pre-storing the fixed matrix inside the analog arrays using 1-bit cells.
Therefore, each bitline corresponds to an output of the \crRW{\texttt{MixColumns}} operation.
Our key insight leverages the observation that although we could \crRW{convert} the full N-bit output from the ADC, only 1-bit is needed due to the subsequent \crRW{\texttt{XOR}}.
Specifically, this implies that the ADC can be limited to 2-bits (e.g., SAR ADC), or can terminate after iterating through the 2-bits (e.g., ramp ADC), saving both energy and time.

\section{Methodology}
\label{sec:methodology}

We model our system using a heavily modified version of MASTODON~\cite{mastodon.github}, a \revsg{cross-stack} simulator for \crRW{digital} PUM.
We integrate MASTODON with CrossSim~\cite{Feinberg.crosssim}, a high-fidelity analog crossbar simulator, which \revsg{faithfully models peripherals (e.g., ADCs, DACs) as well as analog non-idealities.
We use the MILO model~\cite{Milo2021.irps} with CrossSim to capture programming noise and parasitics.}
\revsg{We model a DARTH-PUM chip running at \SI{1}{\giga\hertz}, based on prior SPICE-based circuit simulations.
Table~\ref{tab:methodology:HCT} shows the configuration of each HCT in DARTH-PUM.}

\begin{table}[h]
\vspace{-5pt}
\caption{Hybrid compute tile configuration.}
\label{tab:methodology:HCT}
\centering
\small
\setlength\tabcolsep{3pt}
\begin{tabularx}{\columnwidth}{|l|X|}
\hline
\multicolumn{2}{|c|}{\textbf{\revRW{1 Digital Compute Element}}} \\
\hline
\textbf{Number of Pipelines} & 64 \\
\textbf{Pipeline Depth} & 64~arrays  \\
\textbf{ReRAM Array Size} & 64 $\times$ 64 \\
\hline\hline
\multicolumn{2}{|c|}{\textbf{\revRW{1 Analog Compute Element}}} \\
\hline
\textbf{Number of Arrays} & 64 \\
\textbf{ReRAM Array Size} & 64 $\times$ 64  \\
\textbf{Number of ADCs} & SAR: 2; Ramp: 1 \\
\textbf{ADC Latency} & SAR: 1~cycle; Ramp: 256~cycle \\
\hline
\end{tabularx}
\vspace{-5pt}
\end{table}

\revsg{We compare DARTH-PUM to a Baseline configuration that contains an Intel Core i7-13700~\cite{i13700} CPU and a \SI{1.5}{\giga\byte} ReRAM analog PUM accelerator.
We perform iso-area comparisons, where DARTH-PUM fits in the area of only the CPU (\SI{2.57}{\centi\meter}$^2$), with all hardware parameters scaled to \SI{15}{\nano\meter} to be consistent with prior work~\cite{Stillmaker2017.integration, Truong2021.micro, Oliveira2024.hpca}.
Table~\ref{tab:methodology:area} shows the area breakdown for a DARTH-PUM HCT.
We derive our ADC parameters from prior work on ramp ADCs~\cite{Xiao2022.tcas} and SAR ADCs~\cite{Kull2013.jssc}\crRW{, and incorporate a modified front-end module from the MPU~\cite{Truong2026.hpca}}.
With SAR ADCs, an iso-area DARTH-PUM chip contains \crRW{1860}~HCTs (total memory capacity: \crRW{\SI{4.1}{\giga\byte}}), while with ramp ADCs it contains \crRW{1660}~HCTs (\crRW{\SI{3.7}{\giga\byte}}).}

\begin{table}[h]
\vspace{-5pt}
\caption{\revRW{Area and power for HCT hardware.}}
\label{tab:methodology:area}
\centering
\small
\begin{tabular}{|c|c|c|c|}
\hline
\textbf{DCE} & Area (\SI{}{\micro\meter}$^{2}$) & \textbf{ACE} & Area (\SI{}{\micro\meter}$^{2}$)\\
\hline
ReRAM Array & 240 & ReRAM Array & 240 \\
Pipeline Control & 74000 & Input Buffers & 27000 \\
IO Ctrl & 9600 & Row Periphery & 13000 \\
Decode \& Drive & 280 & \revRW{SAR/RAMP ADC} & \revRW{600/3800} \\
Pipeline Select & 64 & Sample \& Hold & 62 \\
\hline
\hline
\textbf{\crRW{HCT}} & Area (\SI{}{\micro\meter}$^{2}$) & \textbf{\crRW{HCT}} & Area (\SI{}{\micro\meter}$^{2}$)\\
\hline
Shift Unit & 946 & Transpose Unit & 1760 \\
A/D Arbiter & 0.6 & Instr. Injection Unit & 42 \\
\hline
\multicolumn{3}{|c|}{\crsg{Front End (shared by 8 HCTs; \SI{63}{\milli\watt} power)}} & \crsg{87000} \\
\hline
\hline
\textbf{Component} & Power (\SI{}{\milli\watt}) & \textbf{Component} & Power (\SI{}{\milli\watt})\\
\hline
Array (Bool Ops) & 8 &  Row Periphery &  0.7 \\
Pipeline Ctrl & 1.6 & SAR ADC & 1.5 \\
S\&H (Analog) & 2.1$\times$10${^{-5}}$ & Ramp ADC & 1.2 \\
\hline
\end{tabular}
\vspace{-5pt}
\end{table}

To demonstrate the flexibility of DARTH-PUM, we \revsg{evaluate it with} three applications:
(1)~AES~\cite{Nechvatal2001.nist} encrypt, %
(2)~ResNet-20~\cite{He2016.cvpr, Reddi2020.isca} \revRW{inference with the CIFAR-10 dataset~\cite{krizhevsky2009.cifar}}, %
(3)~an LLM encoder~\cite{Vaswani2017.nips}.
\revRW{
We use PyTorch~\cite{Paszke2019.neurips, Feinberg.crosssim} implementations of ResNet-20 and the LLM encoder, \crsg{and use} open-source implementations for AES on CPUs~\cite{OpenSSL} and GPUs~\cite{benlwk.aes}.}

\revsg{We compare DARTH-PUM to four architectures in our evaluation:
(1)~\emph{Baseline}, our Intel CPU with an analog PUM accelerator;
(2)~\emph{DigitalPUM}, an iso-area \crRW{\SI{5.3}{\giga\byte}} RACER~\cite{Truong2021.micro, Truong2022.jetcas} chip \crRW{with one \crsg{front end} per 8~clusters} (running at \SI{1}{\giga\hertz} using the OSCAR logic family~\cite{Truong2022.jetcas}, with two pipelines active per cluster to stay within thermal limits);
(3)~\emph{AppAccel}, which compares against different application-specific accelerators; and
(4)~\emph{GPU}, an NVIDIA GeForce RTX 4090~\cite{RTX4090.nvidia}.
CPU and GPU runs are performed on real hardware, with data collected using performance counters.}

\revsg{AppAccel represents a different design for each application.
For AES, AppAccel uses Intel's AES-NI~\cite{aes-ni} accelerator.
For ResNet-20, AppAccel uses a ReRAM version of a state-of-the-art CNN accelerator~\cite{Xiao2022.tcas}, which uses a ramp ADC and current integrators for the shift-and-add and peripheral ALUs.
For LLM encoding, AppAccel is modeled after an ISAAC-style accelerator~\cite{Shafiee2016.isca}, 
with SAR ADCs and an SFU from recent work~\cite{Song2025.isca} that implements all supplemental functionality (e.g., shift, add, sqrt, ReLU, layernorm).}

\section{Evaluation}
\label{sec:eval}

\revsg{For most of the evaluation, we compare \emph{DARTH-PUM} to \emph{Baseline}, \emph{DigitalPUM}, and \emph{AppAccel}, iso-area to the Baseline CPU.
We use SAR ADCs for all four configurations as it leads to overall improved performance; we show the effect of using ramp ADCs in Section~\ref{ssec:eval:adcs}.
We conduct iso-area comparisons with \emph{GPU} in Section~\ref{sec:eval:gpu}.}

\subsection{\revsg{Performance}}
\label{sec:eval:perf}

\revsg{Figure~\ref{fig:eval:topline} shows the iso-area throughput of each configuration
across the three workloads that we examine.
We make four observations from the figure.
First, we observe that DARTH-PUM provides an average throughput improvement of \crRW{31.4$\times$} over Baseline, with sizable benefits for all three applications over both Baseline and DigitalPUM.
This illustrates DARTH-PUM's ability to adapt across a variety of applications.}

\begin{figure}[h]
    \centering
    \vspace{-5pt}
    \includegraphics[width=0.95\linewidth]{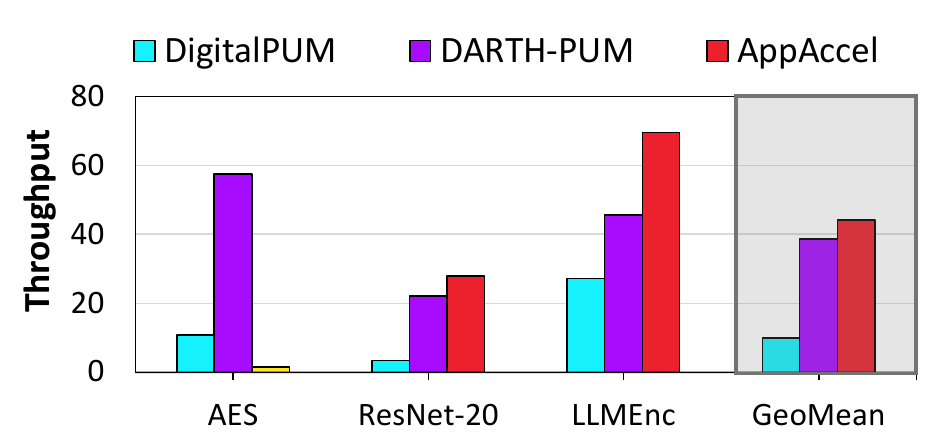}
    \caption{\revsg{Throughput normalized to Baseline. \crRW{AppAccel-AES uses an AES-NI accelerator (yellow), \crsg{while the other AppAccel configurations use analog PUM accelerators (red)}.}}}
    \label{fig:eval:topline}
\end{figure}

\revsg{Second, we observe that \revsg{for AES,} DARTH-PUM outperforms AppAccel by \crRW{36.9}$\times$.
To investigate the source of these benefits, Figure~\ref{fig:eval:aes} breaks down the execution time spent by each configuration on the individual kernels of the AES algorithm.
We find that the latency of a single encryption in DARTH-PUM improves by \crRW{53.7\%} over Baseline.
These savings predominantly come from DARTH-PUM's ability to 
(1)~reduce inter-kernel data movement; and
(2)~employ analog PUM to significantly reduce the \crRW{\texttt{MixColumns}} kernel latency, for which it provides an \crRW{11.5}$\times$ improvement over DigitalPUM.
Combined with Figure~\ref{fig:eval:topline}, we conclude that DARTH-PUM's throughput benefits over Baseline come from a combination of reduced execution time and increased parallelism.}

\begin{figure}[t]
    \centering
    \includegraphics[width=0.95\linewidth]{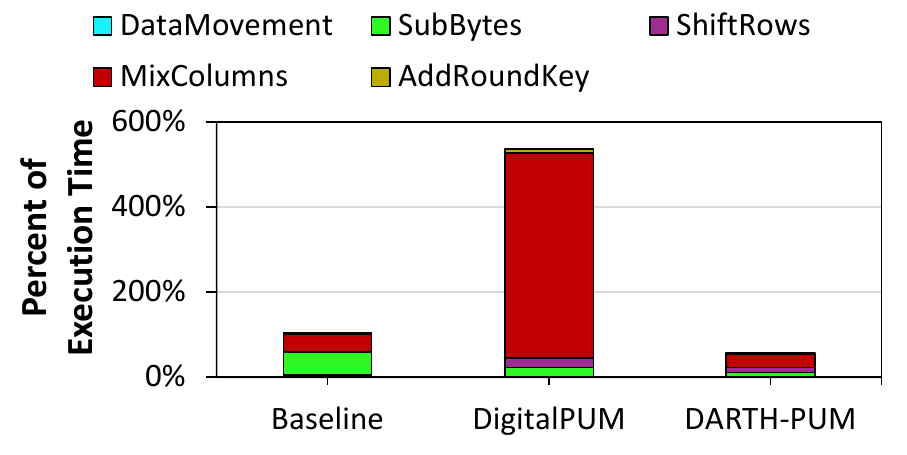}
    \caption{\revsg{Kernel latency breakdown for AES, normalized to Baseline.}}
    \label{fig:eval:aes}
\end{figure}

\revsg{Third, we observe that DARTH-PUM approaches \revsg{AppAccel's throughput} for ResNet-20, coming within \crRW{26.2\%}.
As we can see from Figure~\ref{fig:eval:resnet-20}, similar to AES, DARTH-PUM's improvement over Baseline throughput is not just a result of improved inference latency (which decreases by \crRW{40.0}\%).
In the case of ResNet-20, AppAccel sacrifices a significant amount of its chip area for the SFUs needed to compute each of the model layers, reducing the amount of parallel inference calculations it can perform in an iso-area comparison.
DARTH-PUM's HCT design is significantly more area-efficient than AppAccel's analog PUM plus SFU design, allowing it to make up the throughput gap.}

\begin{figure}[t]
    \centering
    \includegraphics[width=\linewidth]{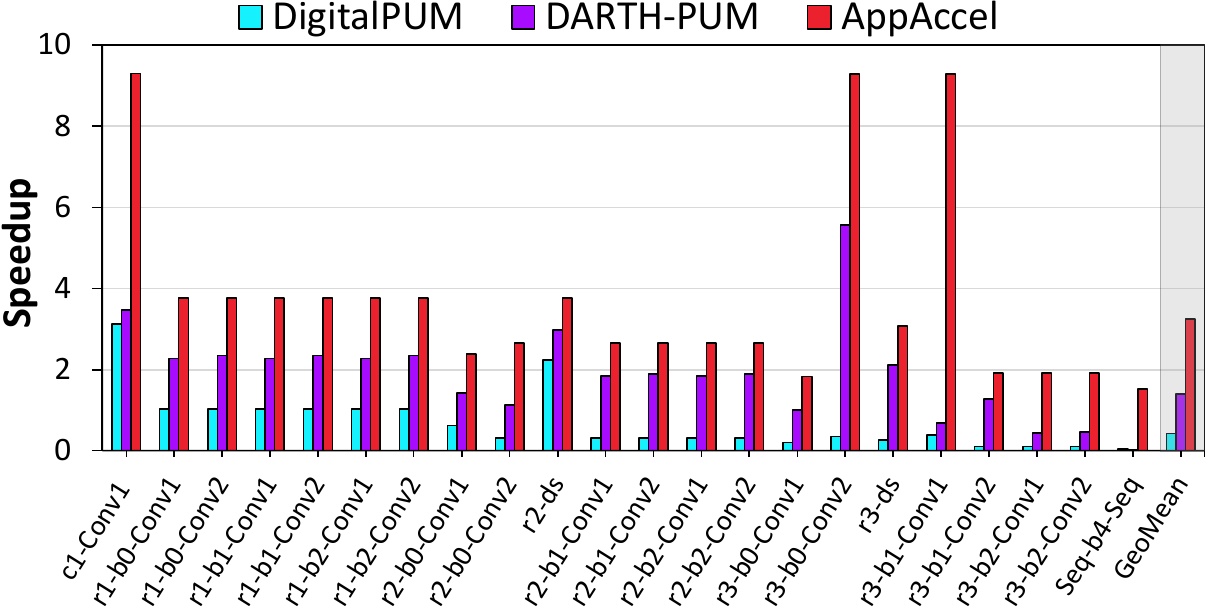}
    \caption{Per-layer \crRW{speedup} for ResNet-20 \crRW{during inference, normalized to Baseline.}}
    \label{fig:eval:resnet-20}
\end{figure}

\revsg{Fourth, we observe that \revad{while }DARTH-PUM still falls short of AppAccel for LLMEnc, it significantly bridges the gap compared to other prior architectures.
The main reason that DARTH-PUM falls short is because 71\% of the total execution time is spent on non-MVM operations (e.g., softmax, ReLU, layernorm), which AppAccel performs with multiple highly specialized SFUs.
Compared to Baseline, 
DARTH-PUM achieves a \crRW{45.6$\times$} benefit.}

\revsg{We conclude that DARTH-PUM is effective at providing large performance benefits over other multi-domain architectures, and captures many of the benefits of application-specific architectures without the need for any SFUs.}

\subsection{Energy}
Processing data locally using \revRW{DARTH-PUM} can provide substantial energy savings \revRW{compared to Baseline, which relies on the CPU to do non-matrix operations.}
Figure~\ref{fig:results:energy} shows the normalized energy required to execute each application.
\revRW{From the figure, we observe that DARTH-PUM achieves an average savings of \crRW{66.8$\times$} compared to Baseline.}
\revRW{When we compare DARTH-PUM \revsg{to DigitalPUM, we observe energy savings of \crRW{2.0$\times$}.}
Notably, this reduction in energy comes from the substantial reduction in digital Boolean operations required to execute the MVM and matrix multiply operations.
\revsg{Across the three applications, we observe that DARTH-PUM achieves competitive energy savings with AppAccel, though its biggest shortfall is for ResNet-20, due to its MVM-heavy nature.}}

\begin{figure}[t]
    \centering
    \includegraphics[width=0.93\linewidth]{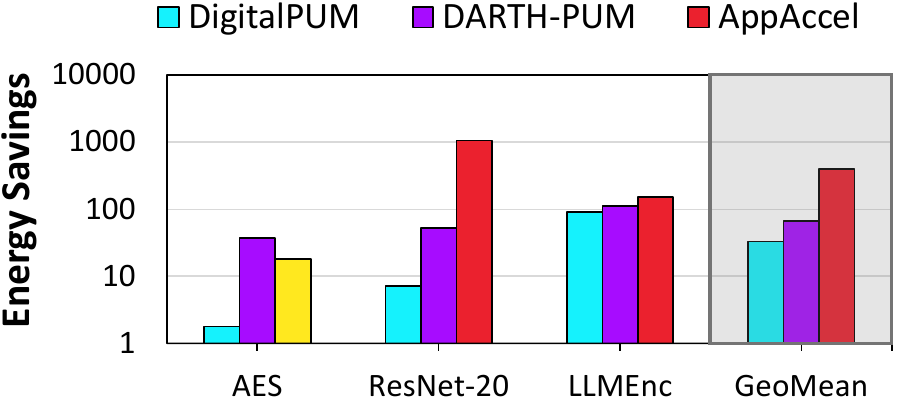}
    \caption{Energy savings\crRW{, normalized to Baseline} \revsg{(y-axis in log scale)}. \crRW{AppAccel-AES uses an AES-NI accelerator (yellow), \crsg{while the other AppAccel configurations use state-of-the-art analog PUM accelerators (red)}.}}
    \label{fig:results:energy}
\end{figure}

\revsg{Overall, we conclude that DARTH-PUM delivers its throughput improvements with significant energy savings across our mix of applications.}

\subsection{\revsg{Impact of Analog-to-Digital Converter Choice}}
\label{ssec:eval:adcs}

\revsg{In this section, we explore whether using ramp ADCs instead of SAR ADCs produces any benefit for those architectures that include analog PUM (see Section~\ref{sec:system:interface}).
Figures~\ref{fig:eval:adc:perf} and \ref{fig:eval:adc:energy} show the performance and energy savings, respectively, 
with both SAR and ramp ADCs.
We make two observations from the figure.
First, for DARTH-PUM, we find that SAR ADCs outperform ramp ADCs by \crRW{1.5$\times$}, while achieving \crRW{99\%} of the energy savings.
This is because across our applications, \crRW{Boolean} PUM \crRW{operations are} responsible for over \crRW{88.1}\% of the total energy consumption \crRW{(9.4\% is consumed by the \crsg{front-end} unit)}, minimizing the ADC's impact on energy.
The application that benefits most from using SAR ADCs is ResNet-20, due to its extensive use of analog PUM.
Second, the only application for which ramp ADCs outperform SAR ADCs is AES, due to an additional optimization opportunity in the \crRW{\texttt{MixColumns}} kernel.
This is due to two reasons:
(1)~we can terminate the ramp ADCs early because we need to check only \revRW{four} bit states (bringing its latency down from 256~cycles to \revRW{four}); and
(2)~while the ramp ADCs can convert all 64 bitlines concurrently, the SAR ADCs are multiplexed across multiple bitlines, reducing their throughput.}

\begin{figure}[b]%
    \centering%
    \includegraphics[width=0.95\linewidth]{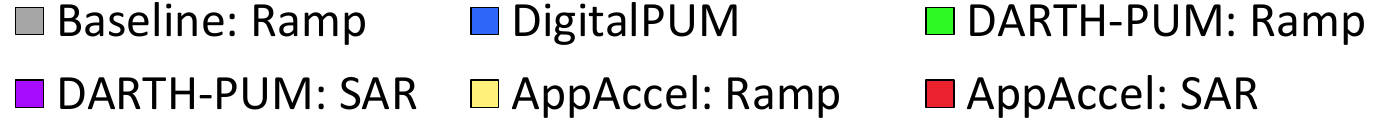}\\%
    \vspace{-2pt}%
    \subfloat[\crRW{Throughput}]{\includegraphics[width=0.48\linewidth]{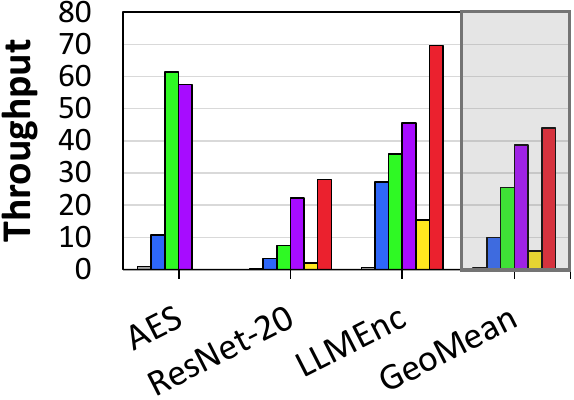}%
    \label{fig:eval:adc:perf}%
    }%
    \hfill%
    \subfloat[\crRW{Energy savings (y-axis in log)}]{\includegraphics[width=0.48\linewidth]{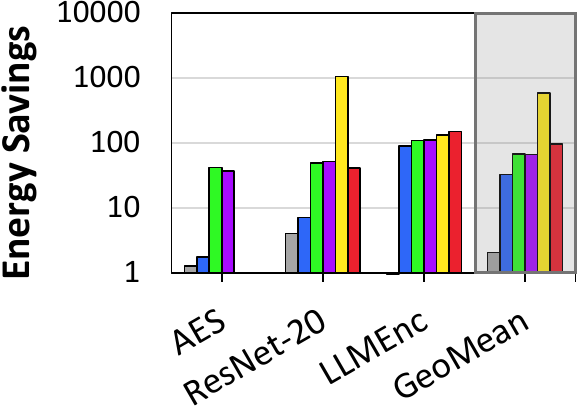}%
    \label{fig:eval:adc:energy}%
    }%
    \caption{\revsg{Comparison of SAR ADCs vs.\ ramp ADCs\crRW{, normalized to Baseline: SAR.}}}%
    \label{fig:eval:adc}%
\end{figure}

\subsection{\revRW{\revsg{Iso-Area} Comparison \revsg{With} GPU}}
\label{sec:eval:gpu}

\revRW{
Figure~\ref{fig:eval:gpu} shows an iso-area comparison between DARTH-PUM and \crsg{the NVIDIA GeForce RTX 4090~\cite{RTX4090.nvidia},} \revsg{a state-of-the-art} GPU.
From the figure we make \revsg{three}
observations.
First, \revsg{on average,} DARTH-PUM 
\revsg{achieves a \crRW{11.8$\times$} and \crRW{7.5$\times$} throughput and energy improvement, respectively, over the GPU.
Second, DARTH-PUM's improvements for AES are significantly lower than those over Baseline, because the AES lookup tables are small enough to be cache-resident in the GPU, enabling it to achieve high throughput.
Third, LLM encoding observes a very large energy savings over the GPU with both DARTH-PUM and DigitalPUM, because it can accelerate the full encoder operations using PUM.}
\revsg{We} conclude that DARTH-PUM 
\revsg{offers promising benefits over GPUs across our range of applications.}}

\begin{figure}[h]%
    \subfloat[\crRW{Throughput}]{\includegraphics[width=0.47\linewidth]{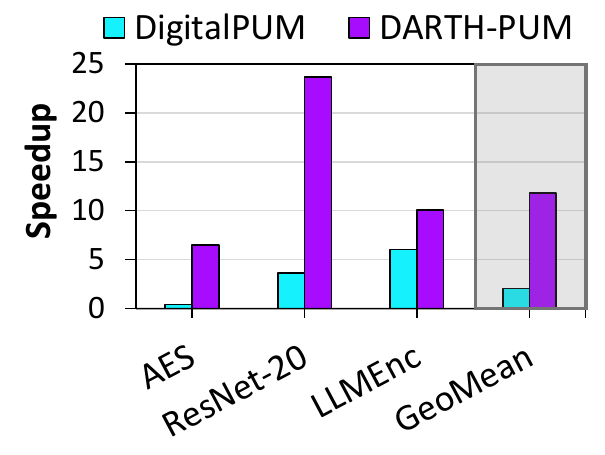}%
    \label{fig:eval:gpu:perf}%
    }
    \hfill%
    \subfloat[\crRW{Energy savings}]{\includegraphics[width=0.49\linewidth]{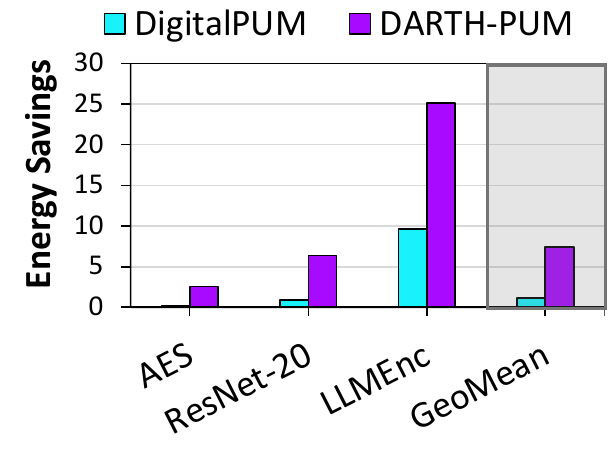}%
    \label{fig:eval:gpu:energy}%
    }%
    \caption{\crRW{Iso-area comparisons to state-of-the-art GPU.}}
    \label{fig:eval:gpu}
\end{figure}

\subsection{\revsg{Impact of Device Non-Idealities \& Noise}}
\label{ssec:eval:accuracy}

\revRW{
\revsg{Because of its use of many conductance levels per cell, analog computation in memory is highly susceptible to inherent noise and non-idealities that occur in many memory devices.
Key sources of these errors are programming noise, various device parasitics (e.g., IR drop), read noise (i.e., error sources during read/MVM operations),
cell \revRW{drift} (i.e., the device conductance can change over time, and for a few device technologies, at random),
``stuck-at'' faults (i.e., devices may become stuck at a fixed conductance state and cannot be overwritten correctly), and manufacturing process variation.\footnote{\revsg{For an in-depth discussion on noise and mitigation techniques, we refer the reader to prior work~\cite{Xiao2020.apr, Xiao2023.cas}}.}
While CrossSim models the first two error sources in detail, all five can contribute to errors in DARTH-PUM.}}

\revsg{Various mitigation techniques, including previously proposed ones for analog PUM and our parasitic compensation scheme (Section~\ref{ssec:system:parasitics}), can address these errors sufficiently.
For example, recent works have demonstrated how to increase ReRAM programming precision~\cite{Song2025.science},
how to partially mitigate read noise using input bit-slicing~\cite{Xiao2023.cas} (which we incorporate in DARTH-PUM), and how to use digital-style error-correcting codes (ECC) to mitigate ``stuck-at'' faults~\cite{Liu2018.itc}.
Even without these techniques, for ResNet-20 using the CIFAR-10 dataset~\cite{krizhevsky2009.cifar}, we observe an end-to-end accuracy of 75.4\%, matching the accuracy of Baseline and AppAccel.
Similar to prior work~\cite{Jiang2020.iscas, Rashed2021.iccad, Lammie2025.jetc}, for use cases that need very high accuracy, select MVM operations can be migrated to \crRW{digital} PUM.
To faithfully report a comprehensive reliability for DARTH-PUM requires detailed chip-level variation/noise models and extensive datasets for device fabrication metrology, which we leave for future work.}

\section{Related Works}
\label{sec:related}

\paragraph{Hybrid Analog--Digital PUM}
Prior works have
proposed hybrid analog--digital PUM as app\-li\-ca\-tion-spe\-ci\-fic accelerators.
Jin et al.~\cite{Jin2022.tpds} propose a hybrid system for training CNNs.
Similarly, Liu et al.~\cite{Liu2025.tcad} use hybrid PUM for hyperdimensional computing.
\revRW{Even though some digital operations} are realized in PUM, \revRW{many of these proposals} still use additional peripherals (e.g., ALUs, \revsg{SFUs}).
\revRW{Works such as SPRINT~\cite{Yazdanbakhsh2022.micro} focus on accelerating the attention mechanism inside the encoder, but leave opportunities to improve the feed-forward \crRW{network}.}
DARTH-PUM uses high-performance digital PUM pipelines to realize non-MVM operations and balances data transfer and processing between analog and digital PUM.

\revRW{
Other works~\cite{Rashed2021.iccad, Jiang2020.iscas} \revsg{attempt} to increase reliability by decomposing applications or operands into analog or digital PUM based on their importance in the overall computation.
This remapping can be done at the application level~\cite{Chen2024.aspdac, Song2025.isca} wherein low-precision analog PUM is used for computations less critical to end-to-end accuracy, and high-precision digital PUM is used for computations more critical to the computation.
DARTH-PUM is amenable to \revsg{such} optimizations by carefully mapping the data into \revsg{the ACE and DCE}.}

\revRW{%
\crsg{Recent works mix}
multiple memory device technologies~\cite{Krishnan2022.tcad, Bhattacharjee2023.jetcas, Zhang2024a.dac, Zhang2024b.dac, Xu2024.dac, Wang2025.tvlsi} for accelerating specific applications.
\crsg{These approaches introduce integration complexities that can significantly raise fabrication prices and hurt yield.}
DARTH-PUM uses a single memory device technology \revsg{to improve manufacturability,} while \revsg{benefiting} multiple applications with the same chip.}

\paragraph{Analog Accuracy}
\revsg{Non-idealities present} a reliability problem for systems leveraging analog MVM.
Therefore, \revsg{many} prior works have examined improving the reliability of analog MVM for different applications, from ML~\cite{Feinberg2018.hpca, Xiao2023.cas, Andrulis2023.isca, Song2025.isca, Lammie2025.jetc, Wang2025.dac, Hou2025.date, Hou2025.iccad} to scientific computing~\cite{Feinberg2018.isca, Feinberg2021.hpca, Song2023.sc}.
\revsg{DARTH-PUM} is amenable to these prior works, as it can support different data layouts or additional circuitry for reliability.
\revRW{Furthermore, this work develops a novel parasitic compensation scheme that leverages both analog and digital PUM to achieve high performance \revsg{with low error counts}.}

\paragraph{\revRW{Analog CAMs}}
\label{para:related:analogcam}
\revRW{
Beyond MVM, prior work~\cite{Li2020.naturecomm, Pedretti2021, Pedretti2022.iscas, Zhao2025.arxiv, Zhao2025.iccd} has proposed \revsg{to use} analog devices for content-addressable memories (CAMs).
At their core, these devices operate similar to  \revsg{SRAM-based} CAMs, \crsg{though} analog CAMs are capable of multi-bit matching,
increasing both their throughput and densities.
Therefore, these devices have been applied workloads including tree-based machine learning~\cite{Pedretti2021, Pedretti2022.iscas}, CNNs and DNNs~\cite{, Zhao2025.arxiv}, transformers~\cite{Zhao2025.iccd}, and network routing~\cite{Li2020.naturecomm}.
Although promising, these devices are susceptible to noise, limiting their \revsg{generalizability} to noise-susceptible applications (e.g., AES), and are unable to handle control flow, both of which are achieved by DARTH-PUM.}

\paragraph{\revRW{ADC-Less Analog}}
\revRW{
The overheads associated with analog periphery can reduce the overall benefit of analog MVM accelerators.
Therefore, prior work~\cite{Negi2024.tcasai, Negi2025.aspdac} has attempted to mitigate these overheads by co-designing the hardware and algorithm to use lower-precision ADCs~\cite{Huang2021.aspdac, Kim2022.jetc}, or in some cases removing them altogether~\cite{Yan2022.isscc, Saxena2023.ispled, Kim2023.jssc, Li2023.tcasii, Chen2023.aspdac, Negi2024.tcasai, Negi2025.aspdac} 
However, these proposals often rely on specific functions known a priori, or assumptions on the specific application of interest and datasets (e.g., approximators for ReLU in DNNs).
In contrast, to maintain flexibility, DARTH-PUM continues to use conventional ADCs, enabling it to be broadly used \revsg{by a wide} variety of applications.
}

\section{Conclusion}
\label{sec:conclusion}
Both analog and \crRW{digital} processing-using-memory continues to be a promising solution towards alleviating costs of data movement. 
DARTH-PUM provides a complete hybrid analog--digital PUM system with ISA, which can be used to accelerate a variety of applications from machine learning to cryptography.
We demonstrate that DARTH-PUM can provide speedups of \crRW{59.4$\times$}, \crRW{14.8$\times$}, and \crRW{40.8$\times$} over an analog+CPU baseline for AES, ResNet-20, and LLM encoders, respectively.

\begin{acks}
We thank T. Patrick Xiao and \revRW{Aniket Das} for their feedback, suggestions, and support throughout this project. 
\revsg{This work was supported in part by
the Sandia National Laboratories Academic Alliance Program,
the Univ.\ of Illinois Center for Advanced Semiconductor Chips with Accelerated Performance (ASAP; an NSF IUCRC),
the Univ.\ of Illinois DREMES HYBRID Center, and
NSF grant CCF-2329096.}

This article has been authored by an employee of National Technology \& Engineering Solutions of Sandia, LLC under Contract No. DE-NA0003525 with the U.S. Department of Energy (DOE). 
The employee owns all right, title and interest in and to the article and is solely responsible for its contents. 
The United States Government retains and the publisher, by accepting the article for publication, acknowledges that the United States Government retains a non-exclusive, paid-up, irrevocable, world-wide license to publish or reproduce the published form of this article or allow others to do so, for United States Government purposes. 
The DOE will provide public access to these results of federally sponsored research in accordance with the DOE Public Access Plan (\url{https://www.energy.gov/downloads/doe-public-access-plan}). 
\end{acks}

\bibliographystyle{ACM-Reference-Format}
\bibliography{refs}

\end{document}